%% file: sound_speed.tex
\documentclass[showpacs,twocolumn,aps,amssymb]{revtex4}
\usepackage{bm}

\begin{document}

\def\ba{{\bf a}}
\def\bk{{\bf k}}
\def\bp{{\bf p}}
\def\bq{{\bf q}}
\def\br{{\bf r}}
\def\bv{{\bf v}}
\def\bx{{\bf x}}
\def\P{{\bf P}}
\def\bR{{\bf R}}
\def\bK{{\bf K}}
\def\la{\langle}
\def\ra{\rangle}
\def\beq{\begin{equation}}
\def\eeq{\end{equation}}
\def\bea{\begin{eqnarray}}
\def\eea{\end{eqnarray}}
\def\bdm{\begin{displaymath}}
\def\edm{\end{displaymath}}

\title{Bogoliubov sound speed in periodically modulated
 Bose-Einstein condensates}
\author{E. Taylor and E. Zaremba}
\affiliation{Department of Physics, Queen's University, Kingston, Ontario
 K7L 3N6, Canada.}
\date{\today}
\begin{abstract}
We study the Bogoliubov excitations of a Bose-condensed gas in an
optical lattice. Of primary interest is the long wavelength phonon
dispersion for both current-free and current-carrying
condensates. We obtain the dispersion relation by carrying out
a systematic expansion of the Bogoliubov equations in powers of the
phonon wave vector. Our result for the current-carrying case agrees 
with the one recently obtained by means of a hydrodynamic theory.
\end{abstract}
\maketitle

\input{intro.tex}
\input{basic.tex}

\input{no_current.tex}
\input{with_current.tex}
\input{conclusions}
\input{appendix1.tex}

\input{bibliography.tex}
\end{document}

%% file: intro.tex
\section{Introduction}
The possibility of creating optical lattices in trapped Bose-condensed
gases has provided an opportunity to study superfluids in novel 
situations. The presence of the lattice leads to a variety of solid
state effects associated with the coherent motion of the atoms in a 
periodic
potential. For example, the oscillation frequency of the centre of mass
motion of the condensate is reduced~\cite{cataliotti01} as a result of 
the enhanced effective mass of the atoms tunnelling between 
potential wells. Furthermore, when subjected to a uniform force, as
provided by gravity~\cite{anderson98} or alternatively by accelerating 
the optical lattice itself~\cite{morsch01}, Bloch oscillations of the
condensate have been observed. Reducing the amplitude of the lattice
potential leads to a breakdown of these oscillations as a result of
Landau-Zener tunnelling between bands~\cite{morsch01}. All of these
observations are essentially a manifestation of the superfluidity
of the Bose condensate in an optical lattice. Another aspect of
equal interest is the breakdown of superfluidity as recently 
observed~\cite{burger01} in a study of the centre of mass motion
of a trapped condensate moving through an optical lattice. When the
amplitude of the oscillation exceeded a critical value, dissipation was
seen to set in.

In this paper we study a Bose-condensed gas subjected to a uniform
optical lattice in a regime where the dynamics of the condensate
wave function is well described by the time-dependent Gross-Pitaevskii 
(GP) equation. In particular, we are concerned with small amplitude
collective modes which at long wavelength are phonon-like
excitations. The relevant
physical parameters determining the properties of the excitation are
the optical potential amplitude, $V_0$, the lattice constant, $d$, the
mean density, $\bar n$, of the gas, and the magnitude of the
supercurrent.
The problem has been addressed theoretically in a number
of papers using a variety of techniques and approximations. 

Berg-S{\o}rensen and M{\o}lmer~\cite{berg-sorensen98} were the first to
investigate phonon excitations within an optical lattice. They solved
the Bogoliubov equations numerically for a one-dimensional model and
established that the long wavelength excitations are phonon-like, having
an energy dispersion that is linear in the wave vector of the mode. They
also obtained an analytic expression for the sound speed, $s$, which is
based on a combined weak potential and slowly varying approximation.
These calculations were extended by Wu and Niu~\cite{wu01} to the case
in which the condensate carries a current. This work is noteworthy for
having pointed out that the modes exhibit both energetic and dynamic
instabilities for sufficiently large currents. The former instability is
associated with the Landau criterion for the breakdown of superfluidity,
while the latter is related to the onset of dissipation as observed
in~\cite{burger01}. The recent paper by Machholm 
{\it et al.}~\cite{machholm03} explores these instabilities further.
Similar results were obtained by Bronski {\it et al.}~\cite{bronski01}
by considering a special form of the lattice potential, while Konotop 
and Salerno~\cite{konotop02} used a different approach to establish that
the dynamic (or modulational) instability leads to the generation of
solitons.
 
When the potential wells are sufficiently deep, the condensate is 
well-localized on each lattice site and a tight-binding description
becomes useful. Javanainen~\cite{javanainen99} used this picture within
a many-body formulation to derive the phonon dispersion throughout the
Brillouin zone for a one dimensional lattice. This calculation is in
fact equivalent to one based on the Bogoliubov
equations~\cite{chiofalo00} or the discrete version of the
time-dependent GP equation~\cite{smerzi02}. The virtue of these methods
is that they provide analytical expressions for the dispersion relation,
although an accurate {\it a priori} determination of the tight-binding 
parameters involves further numerical calculation. Smerzi
{\it et al.}~\cite{smerzi02} extended the results of 
Javanainen~\cite{javanainen99} by deriving the phonon dispersion for a
current-carrying state and found a dynamic instability that is
responsible for a so-called `superfluid-insulator' transition.

The phonon dispersion at long wavelengths was addressed in
\cite{kramer02} by deriving an energy functional involving
density and phase fluctuations which vary slowly in space. 
The approach is closely allied
to the effective mass approximation used in solid state
physics~\cite{callaway91} and recently applied to Bose gases in optical
lattices~\cite{pu03}, and to multiple-scale
analysis~\cite{konotop02,bender78}. The phonon sound
speed is found to be~\cite{menotti02}
\begin{displaymath}
s = \sqrt{{\bar n \over m^*}{\partial \mu \over \partial \bar n}}
\end{displaymath}
where $m^*$ is an effective mass and $\mu$ is the chemical potential.
The precise definition of the effective mass as $(m^*)^{-1} =
\hbar^{-2}\partial^2 \tilde \epsilon /\partial k^2$, where $\tilde
\epsilon$ is the energy per particle of the $N$-particle system, seems
first to have appeared later~\cite{kramer03,machholm03}. This is an
important point since there are other plausible candidates for the
effective mass. In fact, we shall see that a somewhat different, but
equivalent, definition can be given. In this regard, we note that the
effective mass appearing in both the effective mass~\cite{pu03} and
multiple-scale~\cite{konotop02} theories is that
corresponding to the bare optical potential. In other words, the effect
of the interactions on this parameter is not included, and therefore the
use of the dynamical equations obtained in these theories will not in
general give the correct Bogoliubov sound speed.

Our purpose in this paper is to obtain the long wavelength phonon
dispersion directly from the Bogoliubov equations defining the
collective modes. This is achieved by developing a systematic expansion 
of these equations in powers of the phonon wave vector $q$.
We do this first for the current-free state 
(Sec.~\ref{no_current}), confirming the result for the sound speed 
given above. We then consider
a current-carrying state (Sec.~\ref{with_current})
and obtain the analogous phonon dispersion in this case, reproducing the
result obtained in \cite{machholm03} by means of a hydrodynamic
analysis. Our expansion technique can be viewed as a justification of
the assumptions on which the hydrodynamic approach is based.
Furthermore, it provides explicit perturbative expressions for the
various physical quantities that appear in the theory (for example,
the effective mass).

In Sec.~\ref{basic} we present the theoretical background required for
the calculation of small amplitude collective excitations in an optical
lattice. For the most part we consider a three dimensional optical 
potential with cubic symmetry, although we also touch on systems with
one dimensional modulation as well as radially confined condensates. The
underlying periodicity of the the optical potential implies that the
Bogoliubov equations admit solutions having a Bloch function form. This
aspect accounts for the use of a Bloch function basis in solving these
equations in both the current-free (Sec.~\ref{no_current}) and
current-carrying (Sec.~\ref{with_current}) states. However, different
calculational methods are used in the two cases and these are therefore
presented separately. We also examine various physical limits
(Thomas-Fermi, weak potential, weak coupling and tight-binding) in order
to make contact with previous work. As stated previously, our main
result for the phonon dispersion affirms the result which follows from
the insightful use of hydrodynamic equations to describe the dynamics of
long wavelength fluctuations~\cite{machholm03,kramer02}.

%% file: basic.tex
\section{Basic Theory}
\label{basic}
We consider an extended 3D BEC subjected to standing wave light fields
that give rise to a periodic external potential having
the property $V_{\rm opt}(\br+{\bf R}) = V_{\rm opt}(\br)$, where 
${\bf R}$ is a Bravais lattice vector. For the most part we restrict 
ourselves to a cubic lattice for which ${\bf R} = d(n_1 {\bf \hat x} +
n_2 {\bf \hat y} + n_3 {\bf \hat z})$, with $n_i$ an integer.

We base our analysis on the time-dependent Gross-Pitaevskii (GP) 
equation for the condensate wave function, $\Psi(\br,t)$,
\beq
i\hbar \frac{\partial}{\partial t} \Psi(\br,t) =
\bigg(-\frac{\hbar^2 \nabla^2}{2m} + V_{\rm opt}(\br)
+ g|\Psi(\br,t)|^{2} \bigg)\Psi(\br,t).
\label{TDGP}
\eeq
This equation admits stationary solutions of the form
\bdm
\Psi(\br,t) = \Phi(\br) e^{-i\mu t/\hbar}
\edm
where $\Phi(\br)$ satisfies the time-independent GP equation
\begin{equation}
-{\hbar^2 \over 2m} \nabla^2 \Phi + V_{\rm opt} \Phi + g|\Phi|^2
\Phi = \mu \Phi \label{1}
\end{equation}
with the normalization
\bdm
\int_V d^3r\,|\Phi(\br)|^2 = N,
\edm
where $N$ is the total number of particles in the volume $V$. 
Also of interest is the total energy of the system given by
\beq
E_{\rm tot} = \int_V d^3r\,\Phi^*\left ( -{\hbar^2 \nabla^2 \over 2m} +
V_{\rm opt} \right ) \Phi + {g\over 2} \int_V d^3r\,|\Phi|^4\,.
\label{E_tot}
\eeq
The energy parameter $\mu$ is the chemical potential and is related to
$E_{\rm tot}$ by $\mu = \partial E_{\rm tot}/\partial N$. 

Often the ground state solution of
the GP equation is of interest but we shall also consider states
which have a superfluid flow. These states have a Bloch function form
\bdm
\Phi_{n\bk}(\br) = \sqrt{\bar n} e^{i\bk\cdot \br} w_{n\bk}(\br)
\edm
where $n$ is a band index and $\bk$ is a wave vector restricted to the
first Brillouin zone. 
The factor $\sqrt{\bar n}$, where $\bar n$ is the mean density, is 
introduced in the definition of $w_{n\bk}$ so as to give the
normalization
\begin{equation}
{1\over \Omega} \int_\Omega d^3r |w_{n\bk}(\br)|^2 =1\,,\label{2}
\end{equation}
where $\Omega$ is the Wigner-Seitz volume.
The condensate density is then $n_c(\br)= |\Phi_{n\bk}(\br)|^2= \bar n
|w_{n\bk}(\br)|^2$.
The Bloch function $w_{n\bk}(\br)$ is in general
complex and is the self-consistent solution of
\beq
\left ( -{\hbar^2 (\nabla+i\bk)^2 \over 2m} + V_{\rm opt} + 
g \bar n |w_{n\bk}|^2 \right ) w_{n\bk} = \mu_{n\bk} w_{n\bk}\,.
\eeq
We assume that $w_{n\bk}$ has the
periodicity of the lattice, $w_{n\bk}(\br+{\bf R}) = w_{n\bk}(\br)$,
although it should be noted that period-doubled states also
exist~\cite{machholm03b}. The chemical
potential, $\mu_{n\bk}(\bar n)$, is implicitly a function of the mean 
density and depends on the particular Bloch state being considered.

The superfluid current density in this state is
\bea
{\bf j}_s(\br) &=& {\hbar \over 2mi} ( \Phi_{n\bk}^* \nabla \Phi_{n\bk} -
\nabla \Phi_{n\bk}^*  \Phi_{n\bk} )\nonumber  \\
&=& \frac{\bar n}{m} \hbar \bk |w_{n\bk}(\br)|^2 +
\frac{\hbar \bar n}{2mi} \left [ w_{n\bk}^*\nabla w_{n\bk} - (\nabla
w_{n\bk}^*) w_{n\bk} \right ] \nonumber
\eea
and has the property $\nabla \cdot {\bf j}_s(\br) = 0$.
Introducing the superfluid velocity according to the relation
\beq
{\bf j}_s(\br) = n_c(\br) \bv_s(\br)\,,
\eeq
we have
\beq
\bv_s(\br) = \frac{\hbar}{m} (\bk + \nabla \theta_{n\bk}(\br))
\label{local_v}
\eeq
where $\theta_{n\bk}(\br)$ is the phase of the Bloch function 
$w_{n\bk}$. The spatially-averaged superfluid velocity is
\beq
\langle \bv_s \rangle = {1\over \Omega} \int_\Omega d^3r\, \bv_s(\br) =
\frac{\hbar}{m}(\bk + \langle \nabla \theta_{n\bk} \rangle )\,.
\label{avg_vs}
\eeq
In one dimension, the  periodicity of $w_{nk}(x)$ implies 
$\langle d\theta_{nk}/dx \rangle = 2\pi l/d$, where $l$ is
an integer. By continuity of the phase with $k$, we expect $l$ to have 
a fixed value for a given band, and 
$ \langle v_s \rangle = (\hbar/m)(k+G)$ where $G$ is
some reciprocal lattice vector. For the lowest band, we show in
Appendix~\ref{appendixA} that $G=0$. Thus, 
we arrive at the somewhat surprising conclusion
that $ \langle v_s \rangle = \hbar k/m$. We suspect that similar
results apply in higher dimensions but have not been able to show this
explicitly.

The average superfluid velocity should be distinguished from the 
velocity determining the average current density
\bea
\langle {\bf j}_s \rangle &=& {1\over V} \int_V d^3r  \Phi_{n\bk}^* 
{\hbar \over m i}\nabla \Phi_{n\bk}\nonumber \\
&=& {\bar n \over m\Omega} \int_\Omega d^3r w_{n\bk}^* (\bp +\hbar \bk)
w_{n\bk} \nonumber \\
&\equiv& \bar n \bv_{n\bk}\,.
\label{j_s}
\eea
This velocity is given by~\cite{diakonov02}
\beq
\bv_{n\bk} = {1 \over \hbar} \nabla_{\bk} \tilde \epsilon (\bar n,
\bk)\,,
\eeq
where $\tilde \epsilon (\bar n,\bk)=E_{\rm tot}/N$ is the energy per 
particle in the
state characterized by the mean density $\bar n$ and quasimomentum
$\bk$.

In one dimension, the average current density vanishes at the zone
boundary $k = \pi/d$ if the mean density is below a critical value $\bar
n_c$~\cite{diakonov02}. On the other hand, the average superfluid
velocity is $ \langle v_s \rangle = \hbar \pi /md$. This is not a
contradiction since the local superfluid velocity in (\ref{local_v}) is
averaged differently when calculating the average current density.

Dynamical states of the condensate are determined by the time-dependent
GP equation (\ref{TDGP}). For small amplitude excitations, the 
condensate wave function is expressed as
\beq
\Psi(\br,t) = \left [ \Phi_{n\bk}(\br) + \delta \Phi(\br,t)
\right ] e^{-i\mu_{n\bk} t/\hbar}
\eeq
and the GP equation is expanded to first order in the deviation $\delta
\Phi(\br,t)$. By writing
\beq
\delta \Phi(\br,t) = u_i(\br) e^{-iE_it/\hbar} - v_i^*(\br)
e^{iE_i^* t/\hbar}\,,
\eeq
where $E_i$ is allowed to be complex,
one obtains the following Bogoliubov equations for the quasiparticle 
amplitudes $u_i$ and $v_i$, 
\bea
\hat L u_i(\br) - g\Phi_{n\bk}^2(\br) v_i(\br) &=& E_i u_i(\br)
\nonumber \\
\hat L v_i(\br) - g\Phi_{n\bk}^{*2}(\br) u_i(\br) &=& -E_i v_i(\br)
\eea
where the operator $\hat L$ is defined as
\beq
\hat L \equiv - {\hbar^2 \nabla^2 \over 2m} + V_{\rm opt} +
2g|\Phi_{n\bk}|^2 - \mu_{n\bk}\,.
\eeq
Each distinct solution labelled by the index $i$ corresponds to a
collective excitation of the condensate and $E_i$ represents the 
excitation energy of the mode.
The orthonormality of the quasiparticle amplitudes is specified by
\beq
\int_V d^3r\,\left [ u_i^* u_j - 
v_i^* v_j \right ] = \delta_{ij}\,.
\label{uv_ortho}
\eeq

Since the operator $\hat L$ has the translational symmetry of the
lattice, the Bogoliubov equations admit solutions of the form
\bea
u_i(\br) &=& e^{i(\bq+\bk)\cdot \br} \bar u_i(\br) \nonumber \\
v_i(\br) &=& e^{i(\bq-\bk)\cdot \br} \bar v_i(\br) \nonumber 
\eea
where $\bar u_i(\br)$ and $\bar v_i(\br)$ have the periodicity of the
lattice. These functions satisfy
\bea
\hat L_{\bq,\bk} \bar u_i(\br) - g\bar n w_{n\bk}^2(\br) \bar v_i(\br) 
&=& E_i \bar u_i(\br) \nonumber \\
\hat L_{\bq,-\bk} \bar v_i(\br) - g\bar n w_{n\bk}^{*2}(\br) \bar 
u_i(\br) &=& -E_i \bar v_i(\br)
\label{Bogoliubov}
\eea
with
\beq
\hat L_{\bq,\bk} \equiv - {\hbar^2\over 2m} (\nabla+i\bq+i\bk)^2 +
V_{\rm opt} + 2g\bar n|w_{n\bk}|^2 - \mu_{n\bk}\,.
\label{L_op}
\eeq
Our notation emphasizes that $\bq$ and $\bk$ play distinct roles in the
Bogoliubov equations: the former characterizes the Bloch-like character
of the quasiparticle amplitudes while the latter corresponds to the
quasimomentum of the condensate wave function. 

In the following, we shall also make use of the Hamiltonian
\beq
\hat h_\bk(\bq) \equiv - {\hbar^2\over 2m} (\nabla+i\bq+i\bk)^2+ V_{\rm
opt} + g\bar n|w_{n\bk}|^2 - \mu_{n\bk}
\label{GP_Ham}
\eeq
which for $\bq = 0$ is just the Hamiltonian determining the
time-independent condensate wave function $w_{n\bk}$.

%% file: no_current.tex
\section{Phonon Dispersion for a Stationary Condensate}
\label{no_current}

We begin by considering the simpler situation in which there
is no superfluid flow ($\bk = 0$). In this case, the ground state
solution of the time-independent GP equation can be taken to be real,
and $\bar n w_{n=0,\bk={\bf 0}}^2(\br) = 
\bar n w_{n=0,\bk={\bf 0}}^{*2}(\br) = n_c(\br)$. It
is then convenient to introduce the functions $\psi^\pm_i = \bar u_i \pm
\bar v_i$ and to combine the Bogoliubov equations into a single 
equation for $\psi^+_i$~\cite{hutchinson97},
\begin{equation}
\hat h^2_0\psi^+_i+2gn_c\hat h_0 \psi^+_i = E^2_i \psi^+_i
\label{3}
\end{equation}
where $\hat h_0$ is the Hamiltonian
\begin{equation}
\hat h_0(\bq) = - {\hbar^2 \over 2m} (\nabla +i\bq)^2 +V_{\rm opt} 
+g\bar n |w_{0,{\bf 0}}|^2 -\mu_{0,{\bf 0}}\,, \label{4}
\end{equation}
in which the mean field, $g\bar n |w_{0,{\bf 0}}|^2$, of the
current-free condensate appears.

To solve (\ref{3}), we introduce a complete set of Bloch states which
are solutions of the equation
\begin{equation}
\hat h_0(\bq) w_{n\bq} = \varepsilon_n(\bq) w_{n\bq}\,. \label{5}
\end{equation}
This is a linear Schr\"odinger equation but the solution for $\bq = 0$
and $n=0$ coincides with the self-consistent GP solution 
$w_{0,{\bf 0}}$.
By definition, the band energies, $\varepsilon_n(\bq)$, are referred 
to $\mu_{0,{\bf 0}}$, so that $\varepsilon_0({\bf 0}) = 0$.
The functions $w_{n\bq}$ satisfy the periodicity property 
$w_{n\bq}(\br+{\bf R}) = w_{n\bq}(\br)$ and the orthonormality relation
\begin{equation}
{1\over \Omega}\int_\Omega d^3r \, w^*_{n\bq} w_{n'\bq} =
\delta_{nn'}\,.\label{8}
\label{ortho}
\end{equation}
In addition, at $\bq = {\bf 0}$ they are chosen to be real.

Since $\psi^+_i$ is itself a Bloch function, it can be expanded as
\begin{equation}
\psi^+_i(\br) = \sum_n c_n(\bq) w_{n\bq}(\br)\,.
\label{9}
\end{equation}
The label $i$ represents a band index $m$ and the Bloch wave vector 
$\bq$.  However in the following, we will only be interested in the 
lowest excitation band and will therefore drop the label for 
convenience. Substituting this expansion into (\ref{3}) we obtain
\begin{equation}
\sum_{n'} M_{nn'}(\bq)
\varepsilon_{n'}(\bq) c_{n'}(\bq) = E^2(\bq) c_n(\bq) \label{10}
\end{equation}
where
\beq
M_{nn'}(\bq) = {1\over \Omega}\int_\Omega d^3r\, w^*_{n\bq}(\br)
2gn_c(\br)w_{n'\bq}(\br)+\varepsilon_n(\bq)\delta_{nn'}\,.
\label{11} \eeq
We have
displayed explicitly the $\bq$-dependence of all the variables.

For a cubic lattice, we anticipate a particular eigenvalue $E(q)$ 
which has a linear dispersion of the form
\begin{equation}
E(q) = \hbar s q + \cdots\label{12}
\end{equation}
Our objective is to derive an explicit expression for the
Bogoliubov sound speed $s$. From Eq.~(\ref{3}) it is clear that 
in the $q\to 0$ limit, the eigenvector corresponding to this 
particular eigenvalue will be
\begin{equation}
c_n(0) = \delta_{n0}\,,\label{13}
\end{equation}
where $n=0$ labels the lowest Bloch band, since only this band has a
vanishing energy ($\varepsilon_0(0) = 0$). As a function of $q$, 
the lowest band energy behaves as
\begin{equation}
\varepsilon_0(q) = {\hbar^2 q^2 \over 2m_0}+ {\cal O}(q^4)
\label{14}
\end{equation}
which defines the effective mass $m_0$ of this band. We emphasize that
this band mass is determined by the {\it linear} Schr\"odinger equation
(\ref{5}). More will be said about this later. The phonon
eigenvector is a continuous function of $q$ and, as we shall see,
behaves as $c_n(q) = \delta_{n0} + {\cal O}(q^2)$ for small $q$.

To obtain an expression for $s$ we separate the $n=0$ equation
in (\ref{10})
\bea
&&\hskip -.15truein E^2 c_0(\bq) = M_{00}(\bq)\varepsilon_0(\bq)c_0(\bq)
\nonumber \\
&&\hskip .4truein +\sum_{n'\ne 0} M_{0n'}(\bq) \varepsilon_{n'}(\bq) 
c_{n'}(\bq) \nonumber 
\eea
from the $n \ne 0$ equations
\bea
&&\hskip -.15truein E^2 c_n(\bq) = M_{n0}(\bq)\varepsilon_0(\bq) c_0(\bq) \nonumber \\
&&\hskip .4truein+\sum_{n'\ne 0} M_{nn'}(\bq) \varepsilon_{n'}(\bq)
c_{n'}(\bq)\,.\nonumber
\label{15}
\eea
Since $\varepsilon_0(\bq)$ and $E^2$ are both proportional to $q^2$, 
the latter equation shows that
$c_n(\bq) \propto q^2$ for $n \ne 0$, as claimed.
Thus, to order $q^2$, these equations can be replaced by
\begin{eqnarray}
&&\hskip -.25truein E^2 c_0(0) = M_{00}(0)\varepsilon_0(q)c_0(0)
\nonumber \\
&&\qquad\qquad +\sum_{n'\ne 0} M_{0n'}(0) \varepsilon_{n'}(0) c_{n'}(q)
\nonumber \\
&&\hskip -.25truein 0 = M_{n0}(0)\varepsilon_0(q) c_0(0) 
+\sum_{n'\ne 0} M_{nn'}(0) \varepsilon_{n'}(0)
c_{n'}(q) \label{16}
\end{eqnarray}
Solving for $E^2$, we obtain
\beq
E^2 = \varepsilon_0(q) \Big [ M_{00}
-{\sum_{nn'}}'M_{0n}  ( \widetilde{M}^{-1})_{nn'} M_{n'0} \Big ]
\label{17}
\eeq 
where all quantities within the square brackets are understood to be the
$\bq =0$ values. The prime on the summation indicates that the
terms $n=0$ and $n'=0$ are excluded from the sum. The matrix
$\widetilde{M}_{nn'}$ is the matrix obtained by deleting the first row and first
column of $M_{nn'}$. We note that this combination of matrix elements 
can in fact be written as
\beq
M_{00} -{\sum_{nn'}}'M_{0n} (\widetilde{M}^{-1})_{nn'} M_{n'0} = {1\over
(M^{-1})_{00}}\,.
\label{M_inverse}
\eeq
Thus we find that the square of the sound speed is given by
\beq
s^2 = {1\over 2m_0 (M^{-1})_{00}}\,.
\label{18}
\eeq 

We next relate the sound speed to variations of the chemical potential
with mean density. Writing for simplicity $w_0 \equiv w_{0,{\bf 0}}$ and
$\mu_0 \equiv \mu_{0,{\bf 0}}$, we have
\begin{equation}
-{\hbar^2 \over 2m} \nabla^2 w_0 + V_{\rm opt} w_0 + g\bar n w_0^3
= \mu_0 w_0\,. 
\label{19}
\end{equation}
The derivative of this equation with respect to $\bar n$ is
\begin{equation}
\hat h_0({\bf 0}) w_{0,\bar n} + gw_0^3 + 2g\bar n w_0^2 w_{0,\bar n} = 
\mu_{0,\bar n} w_0
\label{20}
\end{equation}
where we use the notation $(\cdots)_{,\bar n}$ to denote a derivative
with respect to $\bar n$.
Taking the inner product of (\ref{20}) with $w_0$ and noting that
$\hat h_0 w_0 = 0$, we find
\begin{equation}
\mu_{0,\bar n} = {g\over \Omega}\int_\Omega d^3r\, w_0^4 + {2g\bar n
\over \Omega}\int_\Omega d^3r\, w_0^3 w_{0,\bar n}\,. \label{21}
\label{dmu_dn_1}
\end{equation}
To solve (\ref{20}) for $w_{0,\bar n}$, we note that the 
normalization condition in (\ref{8}) implies
\begin{equation}
\int_\Omega d^3r\, w_{0,\bar n}  w_0 = 0\,.\label{22}
\end{equation}
Thus, $w_{0,\bar n}$ is orthogonal to $w_0$ and has the expansion
\begin{equation}
w_{0,\bar n} = \sum_{n\ne 0} a_n w_n \label{23}
\end{equation}
in terms of the (real) $\bq =0$ Bloch functions $w_n \equiv w_{n,
\bq=0}$. Substituting this expansion in (\ref{20}) yields
\begin{equation}
{\sum_{n'}}' M_{nn'} a_{n'} = -{1\over 2\bar n} M_{n0}\,. \label{24}
\end{equation}
Using the expansion (\ref{23}) for $w_{0,\bar n}$ in (\ref{21}) with the
expansion coefficients defined by (\ref{24}), we find
\begin{equation}
\mu_{0,\bar n} = {1\over 2\bar n}\left ( M_{00} - {\sum_{nn'}}' M_{0n}
\left (\widetilde{M}^{-1} \right )_{nn'} M_{n'0}
\right ) \,.
\label{25}
\end{equation}
Comparing this with (\ref{M_inverse}), we see that (\ref{18}) is
equivalent to
 \beq s =
\sqrt{\bar n \mu_{0,\bar n} \over m_0}=\sqrt{{\bar n \over m_0}{\partial
\mu_{0,{\bf 0}}
\over \partial \bar n}} \label{26}\eeq 
This result for an optical lattice was first given by Menotti {\it et
al.}~\cite{menotti02} on the basis of general dynamical considerations.
We see here that it follows directly from the Bogoliubov equations and
also applies in the case of a 3D lattice with cubic symmetry. The
small-$\bq$ expansion can be viewed as a systematic way of implementing
the slowly varying ansatz used by Kr\"amer {\it et al.}~\cite{kramer02}.
The expression for $s$ has the same form
as for a homogeneous gas, with $m_0$ replacing the bare mass $m$ and 
the the density derivative of the chemical potential, $\mu_{0,\bar n}$,
replacing the interaction parameter $g$.  In other
words, at long wavelengths the condensate behaves as a gas of particles
of mass $m_0$ with a compressibility, $\kappa$, given by $\kappa^{-1} =
\bar n (\partial \mu_{0,{\bf 0}}/\partial \bar n)$.

\subsection{Thomas-Fermi Limit}
The Thomas-Fermi (TF) approximation is valid when the density varies in
space on a length scale much larger than the local coherence length
$\xi=\sqrt{\hbar^2/2mgn}$. In this situation, the density is
well-approximated by
\beq 
n_0(\br) = {1\over g}(\mu_0 - V_{\rm opt}(\br))
\label{27} \eeq 
except in regions where $V_{\rm opt} \simeq \mu_0$. This does not occur
if $\mu_0 > [V_{\rm opt}]_{\rm max}$, and we then have 
\beq 
\mu_0 = g\bar n + \overline{V}_{\rm opt}\,,
\label{28}
\eeq 
where $\overline{V}_{\rm opt}$ is the mean
value of the optical potential in the unit cell. Thus, $\mu_{0,\bar n} =
g$ as for a homogeneous gas. Since the effective potential, $V_{\rm
opt}+ gn_0$, in the GP equation is a constant in the TF limit, we
would expect the band mass, $m_0$,  to be close to the free particle 
mass, $m$. One can in fact show that the deviation of $m_0$ from $m$ is
proportional to $V_0^2 (\xi/d)^4$, where $V_0$ is the amplitude of 
the potential
modulation. Since we are assuming that $\xi/d \ll 1$, $m_0 \simeq m$
and the TF sound velocity is
\beq s_{\rm TF} \simeq \sqrt{g\bar n \over m}\,,\label{29} \eeq
as for a uniform gas. It should be noted that this result is
valid even when the amplitude of the density modulation is of order the
mean density $\bar n$, provided only that the inequality $\xi/d \ll 1$ 
is everywhere satisfied.

If $\mu_0 < [V_{\rm opt}]_{\rm max}$, the Thomas-Fermi density develops
`holes' in regions where $V_{\rm opt} > \mu_0$. For a one-dimensional
modulation the density is disjoint, as is the case in
two or three dimensions for sufficiently small density. In this
situation long wavelength propagating phonon excitations cannot exist
within the TF approximation since the necessary fluctuations in the
number of atoms from one lattice cell to the next cannot occur. In
reality, the GP density in regions where $V_{\rm opt} > \mu_0$ is small
but finite and phonon-like excitations continue to exist. However, 
increasing the localization of the density in the potential
minima leads to larger effective masses and eventually the sound
speed $s$ tends to zero. This behaviour cannot be described within
the TF approximation.

\subsection{ Weak-Coupling Limit}
The $g$-dependence of $\mu_{0,\bar n}$ appears explicitly in
(\ref{dmu_dn_1}) and
implicitly through the wave function $w_0$ which satisfies (\ref{19}). 
To extract the dependence in the limit $ g\to 0$, we expand the wave
function as $w_0 = w_0^{(0)} + g(\partial w_0/\partial g)_{g=0} +
\cdots$. Since $w_0$ depends on $g$ through the combination $g\bar n$,
we see that $g(\partial w_0/\partial g) = \bar n (\partial w_0/\partial
\bar n)$. Thus, the combination $\bar n w_{0,\bar n}$
appearing in (\ref{dmu_dn_1}) is proportional to $g$ in the small-$g$ 
limit and the second term on the right hand side of (\ref{dmu_dn_1}) 
is of order $g^2$. We then have
\bea
&&\hskip -.5truein\mu_{0,\bar n} = {g\over \Omega} \int_\Omega d^3r
\big (w_0^{(0)}\big )^4 
\nonumber \\ && + {6g^2
\over \Omega} \int_\Omega d^3r \big (w_0^{(0)}\big )^3 (\bar n w_{0,\bar
n}/g)_{g=0}+\cdots
\eea
As discussed in Ref.~\cite{kramer03}, the first term accounts for the 
effect of the
lattice on the compressibility, $\kappa$, which decreases with
increasing localization of the wave function $w_0^{(0)}$. The second
term shows that $\mu_{0,\bar n}$ deviates from a linear dependence on
$g$ as the strength of the interaction is increased. An explicit
expression for this quadratic correction can be obtained from the
equivalent expression for $\mu_{0,\bar n}$ in (\ref{25}) in which the 
two terms respectively correspond to the two integrals in 
(\ref{dmu_dn_1}).
From the definition of the $M_{nn'}$ matrix in (\ref{11}), we see that
\bea
{1\over 2\bar n}M_{00} &=& {g \over \Omega} \int_\Omega d^3r \big ( 
w_0^{(0)} \big )^4\nonumber \\
&&+{4g^2 \over \Omega} \int_\Omega d^3r \big (w_0^{(0)}\big )^3 
(\bar n w_{0,\bar
n}/g)_{g=0}+\cdots\nonumber \\
{1\over 2\bar n}M_{0n} &=& {g \over \Omega} \int_\Omega d^3r \big ( 
w_0^{(0)} \big )^3 w_n^{(0)} + \cdots\nonumber \\
\left ( \widetilde M^{-1} \right )_{nn'} &=& {\delta_{nn'} \over
\varepsilon_n^{(0)}} + {\cal O}(g)\,. \nonumber
\eea
Thus,
\bea
&&\hskip -.5truein\mu_{0,\bar n} = {g\over \Omega} \int_\Omega d^3r
\big (w_0^{(0)}\big )^4 
\nonumber \\ && -  g^2 {\sum_n}' {6\bar n\over
\varepsilon_n^{(0)}} \left | {1\over \Omega}
\int_\Omega d^3r \big (w_0^{(0)}\big )^3 w_n^{(0)} \right |^2+\cdots
\label{dmu_dn}
\eea
Since the excitation energies are positive, we see that $\mu_{0,\bar n}$
has a negative curvature, which agrees with the numerical results in
Ref.~\cite{kramer03}. The interatomic interaction has the effect of 
increasing the width of the wave function which counteracts the 
localizing effect of the lattice potential.

\subsection{ Weak Potential Limit}
\label{weak_potential}

It is also of interest to obtain an expression for the sound speed 
in the case of a weak optical potential
where perturbation theory applies.  For simplicity we consider a
weak one-dimensional periodic potential $V_{\mathrm{opt}} = 
V_{0}\cos(Gz)$, where $G = 2\pi/d$, applied to an otherwise
three-dimensional system. The relevant GP equation is now
one-dimensional,
\beq -\frac{\hbar^2}{2m}{d^2w_{0}\over dz^2} + V_{0}\cos(Gz)w_{0} +
g\bar{n}w_{0}^{3} = \mu_0 w_{0}\,. \label{30} \eeq
In treating the optical potential as a perturbation, we expand the
wave function as $w_0 = w_0^{(0)} + w_0^{(1)} + w_0^{(2)} + \cdots$, and
chemical potential as $\mu_0 = \mu_0^{(0)} + \mu_0^{(1)} + \mu_0^{(2)} 
+ \cdots$, where the superscript here denotes the order in $V_0$. The
properly normalized wave function in the absence of the potential is 
$w_0^{(0)} = 1$ and $\mu_0^{(0)} = g\bar n$. To first order in $V_0$,
$\mu_0^{(1)} = 0$ and
\beq
w_0^{(1)} = - {V_0\over \varepsilon_G^{(0)} + 2g\bar n} \cos(Gz)\,,
\label{w_0^1}
\eeq
where $\varepsilon_G^{(0)} = \hbar^2 G^2/2m$.
In calculating the second order contribution, $\mu_0^{(2)}$, to the
chemical potential,
the second order wave function, $w_0^{(2)}$, need not be determined, but
the normalization condition $\int_{-d/2}^{d/2} dz\, w_0^{(0)} 
w_0^{(2)} = - (1/2)\int_{-d/2}^{d/2} dz \big ( w_0^{(1)} \big)^2$ is
required. Thus,
the chemical potential correct to second order in $V_0$ is found to be
\beq 
\mu_0 =  g\bar{n} -
\frac{\varepsilon_{G}^{\left(0\right)}V_0^2}{2(\varepsilon_{G}^{(0)}
+ 2g\bar{n})^{2}}+\cdots \label{31}\eeq 
Taking the derivative with respect to $\bar{n}$, we have
\beq 
\mu_{0,\bar n} = g +
\frac{2gV_{0}^{2}\varepsilon_{G}^{(0)}}{(\varepsilon_{G}^{(0)} +
2g\bar{n})^{3}} \,.\label{33}\eeq
The weak-coupling limit of this result to lowest order in $g$ is
$\mu_{0,\bar n} = g(1+2(V_0/\varepsilon_G^{(0)})^2)$. This can be shown
to agree with the expansion of the first term in (\ref{dmu_dn}) to 
second order in $V_0$.

To complete the calculation of the sound speed, we require an equivalent
expression for the effective mass. This can be obtained by solving
\bea
&&\hskip -.25truein -\frac{\hbar^{2}}{2m}\left ( {d\over dz} +iq 
\right )^2 w_{0q} + V_{0}\cos(Gz) w_{0q} \nonumber \\
&&\hskip .5truein + (g\bar{n}w_{0}^{2}  - \mu_0) w_{0q}
= \varepsilon_0(q) w_{0q} \,,
\label{34}
\eea
where $w_0$ is the solution of (\ref{30}). Since the correction to the
effective mass is second order in $V_0$, it is sufficient to consider
the mean-field potential $(g\bar{n}w_{0}^{2}  - \mu_0)$ to first order 
in $V_0$. We must then solve
\begin{displaymath}
-\frac{\hbar^{2}}{2m}\left ( {d\over dz} +iq 
\right )^2 w_{0q} + V_{0}^\prime \cos(Gz) w_{0q}
= \varepsilon_0(q) w_{0q} \,, \nonumber
\end{displaymath}
where $V'_0 = V_0 \varepsilon_G^{(0)} /(\varepsilon_G^{(0)} +2g\bar n)$.
Again treating $V'_0$ perturbatively,
the effective mass for the lowest band is found to be
\beq \frac{1}{m_{0}} = \frac{1}{m}\left[1 -
\frac{2V_{0}^{2}}{(\varepsilon_{G}^{(0)} +
2g\bar{n})^{2}}\right]
\label{meff}
\eeq
Inserting (\ref{meff}) and (\ref{33}) into (\ref{26}), and discarding 
the quartic term in $V_{0}$, we arrive at the following expression for
the sound speed
\beq 
s = \sqrt{\frac{g\bar{n}}{m}\left(1 -
\frac{4g\bar n V_{0}^{2}}{(\varepsilon_{G}^{(0)} +
2g\bar{n})^{3}}\right)}\,.
\label{36}
\eeq
When $\varepsilon_{G}^{(0)}\ll g\bar{n}$ (or $\xi/d \ll 1/2\pi$), 
this expression reduces to
\beq 
s = \sqrt{\frac{g\bar{n}}{m}\left(1 -
\frac{V_{0}^{2}}{2\left(g\bar{n}\right)^{2}}\right)} \label{37}
\eeq
in agreement with the approximation to the sound speed obtained
by Berg-S{\o}rensen and M{\o}lmer~\cite{berg-sorensen98}.

To make contact with Ref.~\cite{kramer03}, we introduce the recoil 
energy $E_R =
\hbar^2 \pi^2/2md^2 = \varepsilon_G^{(0)}/4$ and define $2V_0 =
\sigma E_R$ (the parameter $\sigma$ is called `$s$' in
Ref.~\cite{kramer03}).
Eq.~(\ref{36}) can then be rewritten as
\beq
{s\over s_0} = 1 - {\gamma \sigma^2 \over 128(1+\gamma/2)^3}
\eeq
where $s_0 = \sqrt{g\bar n/m}$ is the sound speed for the homogeneous
gas and $\gamma$ is the ratio $g\bar n/E_R=(d/\pi \xi)^2$. 
This shows that $s$ decreases quadratically with the strength of the 
optical potential, which
is consistent with the numerical results in Ref.~\cite{kramer03}. 
This expression
is valid if $2V_0 \ll \varepsilon_G^{(0)}$, or $\sigma \ll 4$,
however, this constitutes a rather limited range of the values of
$\sigma$ of physical interest.

\subsection{Radially Confined Condensates}
As a final application of the results derived in this section we
consider a condensate that is confined in the radial direction. To be
specific, we assume a potential of the form
\beq
V(\br) = V_{\rm opt}(z) + V_\perp(\rho)
\eeq
where $V_\perp(\rho) = m\omega_\perp \rho^2/2$, that is, harmonic
confinement in the radial ($\rho$) direction. The optical potential is
periodic in the axial direction with periodicity $d$. This potential 
approximates the situation of a long cigar-shaped trap with an axial 
standing light wave.

Although the geometry is quite different from that considered earlier,
the previous analysis can be carried over with minor modification.
The ground state GP solution has the property $\Phi_0(\rho,z+d) =
\Phi_0(\rho,z)$, and has the normalization
\beq 
\frac{1}{d}\int_{-d/2}^{d/2}dz\int d^2 r_\perp
|\Phi_0({\mathbf{r}})|^{2} = \bar \lambda\,,\label{38}
\eeq
which defines the mean linear density $\bar \lambda$ along the length of
the condensate. As before, it is convenient to define
$\Phi_0(\br) \equiv \sqrt{\bar \lambda} w_0(\br)$.

The Bogoliubov excitations in the present situation have a Bloch 
wave character along the axis and are obtained from (\ref{3}) with the
Hamiltonian
\beq
\hat h_0(q) = -{\hbar^2 \over 2m} \left [ \nabla_\perp^2 + \left (
{\partial \over \partial z} +iq \right )^2 \right ]
+ V + g\bar \lambda |w_0|^2 -\mu_0\,.
\eeq
The eigenstates of $\hat h_0(q)$ are now labelled by the set of quantum
numbers $\{q,n,m,\nu\}$, where $n$ is a one-dimensional band index, 
$m$ is the azimuthal quantum number associated with the $z$-component 
of angular momentum, and $\nu$ labels the different
radial excitations. Of interest here are axially symmetric solutions
($m=0$) since these have the character of the phonon mode of interest.

The analysis after (\ref{3}) is followed step by step, the only change
being the integration volume used in the normalization of
the states. We thus find that the sound speed is given by 
\beq
 s =
\sqrt{\frac{\bar{\lambda}}{m_{0}}\frac{\partial \mu_0}{\partial 
\bar{\lambda}}}
\label{40} \eeq 
where $m_0$ is the effective mass of the lowest band ($n=0$ and $\nu
=0$).

Eq.~(\ref{40}) is of course valid in the absence of the optical
potential. Treating the condensate in the TF approximation, we find
$\bar \lambda = \pi \mu_0^2/gm\omega^2_\perp$ and $\partial \mu_0
/\partial \bar \lambda = gm\omega^2_\perp/2\pi\mu_0$. This gives a sound
speed
\beq
s=\sqrt{{gn_0(0)\over 2m}}
\label{s_cyl}
\eeq
where $n_0(0)=\mu_0/g$ is the density at the centre of the trap. This 
result was first obtained in Ref.~\cite{zaremba98} using a different
method.
In the weak coupling limit, (\ref{dmu_dn}) applies. In this case,
the condensate wave function is a gaussian,
\begin{displaymath} w_{0}^{(0)} =
\sqrt{\frac{m\omega_{\bot}}{\pi\hbar}}\exp\left(-\frac{m\omega_{\bot}
\rho^2} {2\hbar}\right)\,,
\label{43}
\end{displaymath}
and one again obtains the result in
(\ref{s_cyl}) for the sound speed, as found previously~\cite{jackson98}.

When the optical potential is strong the condensate becomes localized on
each site. In this situation, the tight-binding approximation is a
useful method for dealing with the
system~\cite{javanainen99,chiofalo00,trombettoni01}. Approximating the
condensate wave function as $\Phi(\br) = \sum_i c_i f_i(\br)$ where
$f_i(\br)$ is a function localized on the $i$-th site and normalized to
unity, the energy of the system is given approximately by
\begin{displaymath}
E_{\rm tot} \simeq \sum_i \varepsilon_0|c_i|^2 - t \sum_i (c_i^* c_{i+1} +
c_i c_{i+1}^*) +{1\over 2} \tilde g\sum_i|c_i|^4\,,
\end{displaymath}
where $\varepsilon_0$ is an on-site energy, $t$ is a hopping matrix
element connecting the amplitudes on nearest-neighbour sites and 
$\tilde g$ is
an effective interaction strength. For the ground state, $|c_i|^2 =
\nu$, where $\nu$ is the number of atoms per site. We then have
\begin{displaymath}
E_{\rm tot} = (\varepsilon_0 -2t)N+ {1\over 2} \tilde g \nu N\,.
\end{displaymath}
Assuming that the parameters $\varepsilon_0$, $t$ and $\tilde g$ are
density independent in the extreme tight-binding limit, we have 
$\mu_0 = \partial E_{\rm tot}/\partial N =
(\varepsilon_0-2t)+\tilde g \nu$ and $\partial
\mu_0/\partial \bar \lambda = \tilde g d$. Within the same approximation
the band energy is $\varepsilon(q) = \varepsilon_0 +\tilde g \nu -
2t\cos(qd)$, which gives an effective mass $m_0 = \hbar^2/2td^2$. Thus
the Bogoliubov sound speed from (\ref{40}) is $s_{tb} \simeq 
\sqrt{2\tilde g\nu t d^2/\hbar^2}$,
which is the result obtained by Javanainen~\cite{javanainen99}.

%% file: with_current.tex
\section{Phonon Dispersion for a moving condensate}
\label{with_current}

We turn next to the derivation of the phonon dispersion relation for
the case where the condensate is ``flowing" through the optical
lattice.  We address this problem by directly solving the Bogoliubov
equations in (\ref{Bogoliubov}) 
for a specific condensate wave function $w_{n\bk}$ in the limit of small
wave vectors $\bq$. The structure of these equations is quite different 
when
$\bk \ne 0$ and the method used in the previous section to determine 
the dispersion relation can no longer be applied. In fact, the analysis
is much more intricate as will soon become apparent. The results
we obtain confirm the more intuitive hydrodynamic approach presented
recently~\cite{machholm03}, which describes the dynamics of the system 
in terms of slowly
varying hydrodynamic variables (density and momentum). By including 
small length scale variations, our approach in a sense provides a 
``microscopic" 
derivation of the hydrodynamic equations that one would expect to be
valid in the long wavelength limit.

Since the solution in the small-$\bq$ limit is required, we rewrite
(\ref{Bogoliubov}) so as to display the $\bq$-dependent terms
explicitly:
\bea
&&\left ( \hat h_\bk + {\hbar \bq \over m} \cdot (\bp +\hbar \bk) +
{\hbar^2 q^2 \over 2m} + g\bar n |w_{0\bk}|^2 \right ) \bar u \nonumber
\\
&&\hskip 1.5truein - g\bar n w_{0\bk}^2 \bar v = E \bar u \nonumber \\
&&\left ( \hat h_{-\bk} + {\hbar \bq \over m} \cdot (\bp - \hbar \bk) +
{\hbar^2 q^2 \over 2m} + g\bar n |w_{0\bk}|^2 \right ) \bar v \nonumber
\\
&&\hskip 1.5truein - g\bar n w_{0\bk}^{*2} \bar u = -E \bar v \,,
\label{newBog}
\eea
where $\bp = (\hbar/i)\nabla$ is the momentum operator.
The GP Hamiltonian here is $\hat h_\bk(\bq=0)$ as defined in
(\ref{GP_Ham}).
In these equations, we have adopted the index $n=0$ for the
condensate wave function. We will usually think of this state as the
lowest Bloch state solution of the GP equation, although it in principle
could correspond to an arbitrary excited band. For simplicity we have
dropped the index on the quasiparticle amplitudes $\bar u$ and $\bar v$
and the excitation energy $E$ as we will only be considering the
phonon-like excitation.

It is clear that for $\bq =0$ (\ref{newBog}) admits a solution with 
$\bar u \propto w_{0\bk}$, $\bar v \propto w_{0\bk}^*$ and $E = 0$. 
For finite $\bq$ we seek solutions in the form of an expansion in 
eigenfunctions of $\hat h_{\pm\bk}$, namely,
\bea
\bar u &=& \sum_n a_n(\bq) w_{n\bk}\nonumber \\
\bar v &=& \sum_n b_n(\bq) w_{n,-\bk}
\label{uv_expansions}
\eea
where
\beq
\hat h_\bk w_{n\bk} = \varepsilon_{n\bk} w_{n\bk}
\eeq
According to this definition, $\varepsilon_{0\bk} \equiv 0$. The
functions $w_{n\bk}$ are an orthonormal set with normalization given by
(\ref{8}). Although we use the same notation, it should be noted that
these functions are distinct from those defined in (\ref{5}).
In addition, we may assume $w_{n,-\bk}(\br) = w^*_{n\bk}(\br)$ and
$\varepsilon_{n,-\bk} = \varepsilon_{n\bk}$. 

Substituting these
expansions into (\ref{newBog}), we obtain the matrix equations
\begin{eqnarray}
&&\hskip -.4truein \left(\varepsilon_{n\bk} +  \frac{\hbar^2q^2}{2m}\right)a_{n}
+ \sum_{n'}\frac{\hbar}{m}\bq \cdot \P_{nn'} a_{n'}\nonumber\\
&& + \sum_{n'}\left ( A_{nn'} a_{n'} -
B_{nn'} b_{n'}\right ) = E a_{n},
\label{a_equation}
\end{eqnarray}

\begin{eqnarray}
&&\hskip -.4truein \left(\varepsilon_{n\bk} +
\frac{\hbar^2q^2}{2m}\right)b_{n} - \sum_{n'} \frac{\hbar}{m} \bq\cdot
\P^{*}_{nn'} b_{n'}\nonumber\\
&& + \sum_{n'}\left ( A^{*}_{nn'} b_{n'}-
B^{*}_{nn'} a_{n'}\right ) = -E b_{n},
\label{b_equation}
\end{eqnarray}
where we have defined the matrices
\begin{eqnarray}
A_{nn'}(\bk) &=& \frac{1}{\Omega}\int_\Omega w_{n\bk}^{*}g \bar n |w_{0\bk}|^2w_{n'\bk} d^{3}r,\label{114}\\
B_{nn'}(\bk) &=& \frac{1}{\Omega}\int_\Omega w_{n\bk}^{*} g\bar nw_{0\bk}^{2}w_{n'\bk}^{*} d^{3}r,\label{115}\\
\P_{nn'}(\bk) &=& \frac{1}{\Omega}\int_\Omega
w_{n\bk}^{*}\left(\bp + \hbar
\bk\right)w_{n'\bk} d^{3}r \label{116}.
\end{eqnarray}

Due to the inversion symmetry of the lattice, the Bloch states at the
zone centre ($\bk = 0$) can be chosen to be simultaneous eigenstates of
parity. Although parity is not a good quantum number for states with 
nonzero $\bk$, each band can nevertheless be assigned a parity index 
$\eta_{n} = \pm 1$ such that

\begin{equation}
w_{n,-\bk}(-\br) = \eta_{n}w_{n\bk}(\br)\,, \label{118a}
\end{equation}
This property is proved explicitly in one dimension in Appendix 
\ref{appendixA} and can also be shown to follow to lowest order in 
$\bk$ by means of $\bk\cdot \bp$ perturbation theory. Together with the
conjugation (time-reversal) property $w^*_{n\bk}(\br) =
w_{n,-\bk}(\br)$, we have
\begin{equation}
w^*_{n\bk}(\br) = \eta_{n}w_{n\bk}(-\br)\,. 
\label{conjugation}
\end{equation}
This important property is used throughout the following discussion.
For example, it can be used to show that the matrices in
(\ref{114}-\ref{116}) satisfy the following relations
\bea
&&A^*_{nn'}(\bk) = A_{n'n}(\bk) = A_{nn'}(-\bk) = \eta_n \eta_{n'} 
A_{nn'}(\bk)\nonumber \\
&&B^*_{nn'}(\bk) = B_{nn'}(-\bk) = \eta_n \eta_{n'} B_{nn'}(\bk) =
\eta_n \eta_{n'} B_{n'n}(\bk)
\nonumber \\
&&\P^*_{nn'}(\bk) = \P_{n'n}(\bk) = -\P_{nn'}(-\bk) = \eta_n \eta_{n'} 
\P_{nn'}(\bk)\nonumber \\
\label{relations}
\eea
Note that $A$ and $\P$ are hermitian while $B$ is not. In addition, we
see that $A_{nn'}(0)$ and $B_{nn'}(0)$ are real and nonzero only for
pairs of states having the same parity index, while $\P_{nn'}(0)$ is
purely imaginary and only couples states with opposite parity.

In solving (\ref{a_equation}) and (\ref{b_equation}), it
is convenient to define the following linear combinations: 
$c_{n} = \frac{1}{2}(a_{n} + \eta_{n}b_{n})$ and
$d_{n} = \frac{1}{2}(a_{n} - \eta_{n}b_{n})$.  Introducing these
variables into (\ref{a_equation}) and (\ref{b_equation}), and 
making use of the relations in (\ref{relations}), we obtain the 
equations
\begin{eqnarray}
&&\hskip -.4truein \left(\varepsilon_{n\bk} +
\frac{\hbar^2q^2}{2m}\right)c_{n}
+ \sum_{n'}\left (A_{nn'}-\widetilde B_{nn'} \right )c_{n'}\nonumber \\
&&+ \sum_{n'}\frac{\hbar}{m} \bq \cdot \P_{nn'} d_{n'}
= E d_{n},\label{117}
\end{eqnarray}

\begin{eqnarray}
&&\hskip -.4truein \left(\varepsilon_{n\bk} +
\frac{\hbar^2q^2}{2m}\right)d_{n} + \sum_{n'}\left ( A_{nn'} + 
\widetilde B_{nn'} \right ) d_{n'} \nonumber \\
&&+\sum_{n'} \frac{\hbar}{m} \bq\cdot \P_{nn'} c_{n'} = E c_{n},
\label{118}
\end{eqnarray}
where we have defined the hermitian matrix $\widetilde B_{nn'} =
B_{nn'}\eta_{n'}$. 
As in our earlier analysis, we anticipate that $E(\bq)$ will depend
linearly on $\bq$ for $q \to 0$, but due to the quasimomentum $\bk$ of 
the condensate, it no longer depends simply on the magnitude $q$.
To extract this dependence we systematically expand the coefficients
$c_{n}(\bq)$ and $d_{n}(\bq)$ as a series in powers of $q$.
Specifically, we write
\begin{eqnarray}
c_{n}(\bq) &=& h(\bq)(c_n^{(0)}+c_n^{(1)} + c_n^{(2)}\cdots)\nonumber\\
d_{n}(\bq) &=& h(\bq)(d_n^{(0)}+d_n^{(1)} + d_n^{(2)}\cdots)\nonumber\\
\label{120a}
\end{eqnarray}
where the superscript indicates the order of $\bq$ in the respective
terms (here, order signifies similar powers of the vector magnitude
$q$~\cite{footnote3}). The factor $h(\bq)$ contains the 
nonanalytic behaviour of the coefficients which is required in order 
to satisfy the normalization condition in (\ref{uv_ortho}).
For a homogeneous system, $h(q)
\propto q^{-1/2}$, and we expect a similar dependence in the case of a
lattice. In the following, it is sufficient to note that this factor is
the same for both coefficients and can therefore be ignored in
developing a systematic $q$-expansion.

We noted earlier that $\bar u \propto w_{0\bk}$ and $\bar v \propto
w_{0,-\bk}$ for $\bq \to 0$ which, according to (\ref{uv_expansions}), 
implies that
$a_n(\bq) \propto \delta_{n0}$ and $b_n(\bq) \propto \delta_{n0}$ in
this limit. If the state $w_{0\bk}$ is an even parity state ($\eta_0 =
+1$), as we assume in the following, we must then have $c_n^{(0)} =
\delta_{n0}$ and $d_n^{(0)} = 0$. With this information, (\ref{117}) 
gives to ${\cal O}(q^0)$ the equation
\beq
\varepsilon_{0\bk} + (A_{n0} - \widetilde B_{n0}) = 0
\eeq
which is satisfied since $\varepsilon_{0\bk} = 0$ and $\widetilde 
B_{n0} = A_{n0}$.  To ${\cal O}(q^1)$, (\ref{117}) gives
\beq
\varepsilon_{n\bk}c_n^{(1)} + \sum_{n'} (A_{nn'} - \widetilde{B}_{nn'})
c_{n'}^{(1)} = 0\,.
\label{c^1}
\eeq 
Similarly, (\ref{118}) gives
\beq
\sum_{n'} N_{nn'} d_{n'}^{(1)} = E\delta_{n0} - {\hbar 
\over m} \bq \cdot\P_{n0}\,,
\label{d^1}
\eeq
where we have defined the hermitian matrix
\beq
N_{nn'}(\bk) = A_{nn'}(\bk) + \widetilde B_{nn'}(\bk)+ 
\varepsilon_{n\bk}\delta_{nn'}\,.
\label{N_nn'}
\eeq
Eq.~(\ref{c^1}) is homogeneous and indicates that $c_n^{(1)} \equiv 0$.
On the other hand, (\ref{d^1}) can be solved for $d_n^{(1)}$ in terms of
the unknown excitation energy $E$. To determine the latter, we must
consider (\ref{117}) to ${\cal O}(q^2)$, obtaining
\bea
&&\hskip -.2truein \varepsilon_{n\bk}c_n^{(2)} + {\hbar^2 q^2 \over 2m} \delta_{n0}
+\sum_{n'} (A_{nn'}-\widetilde B_{nn'})c_{n'}^{(2)} \nonumber \\
&&\hskip .75truein + \sum_{n'} {\hbar
\over m} \bq\cdot \P_{nn'} d_{n'}^{(1)} = E d_n^{(1)}\,.
\eea
Setting $n=0$ in this equation, and noting that $\varepsilon_{0\bk}=0$
and that $A_{0n} = \widetilde B_{0n}$, we find
\beq
{\hbar^2 q^2 \over 2m} + \sum_{n} {\hbar \over m} \bq\cdot \P_{0n}
d_n^{(1)} = E d_0^{(1)}\,.
\label{quad}
\eeq
Since $d_n^{(1)}$ is itself linear in $E$ according to (\ref{d^1}), we
see that (\ref{quad}) is implicitly a quadratic equation for $E$ which
can be solved to determine the excitation energy to lowest order in
$\bq$. However, to do so directly would not reveal the interesting
dependences on various physical parameters that in fact emerge. As seen
in the $\bk =0$ analysis, the excitation energy could be related to the
variation of the chemical potential with mean density $\bar n$. This
remains a quantity of interest in the present case, but we must also
consider variations of the chemical potential with $\bk$. We thus turn 
next
to the determination of $\mu_{0\bk,\bar n} \equiv \partial \mu_{0\bk}
/\partial \bar n$ and $\mu_{0\bk,i} \equiv  \partial \mu_{0\bk}/\partial
k_i$.

\subsection{ Determination of $\mu_{0\bk,\bar n}$}

The chemical potential $\mu_{0\bk}$ is determined by the self-consistent
solution of the GP equation, $\hat{h}_{\bk}w_{0\bk} = 0$, with
$\hat{h}_{\bk}$ given by (\ref{GP_Ham}) with $\bq=0$. The variation 
of this equation with respect to $\bar n$ gives
\begin{eqnarray}
&&\hskip -.25truein \hat{h}_{\bk}w_{0\bk,\bar n} + g\left\vert 
w_{0\bk}\right\vert^2w_{0\bk} +g\bar{n}\left(w^{*}_{0\bk,\bar n}w_{0\bk}
+ w_{0\bk}^{*}w_{0\bk,\bar n}\right)w_{0\bk} \nonumber \\
&&\hskip 2truein = \mu_{0\bk,\bar{n}}w_{0\bk}, \label{130}
\end{eqnarray}
\noindent where $w_{0\bk,\bar n} \equiv \partial w_{0\bk}/\partial 
\bar n$. Taking the derivative of the normalization (\ref{ortho})
with respect to $\bar{n}$, and using (\ref{conjugation}), we find that
\begin{eqnarray}
\int_\Omega w_{0\bk}^{*}w_{0\bk,\bar n}d^{3}r = 0.  \label{131}
\end{eqnarray}
Thus, $w_{0\bk,\bar n}$ is orthogonal to $w_{0\bk}$ and has the
expansion
\begin{eqnarray}
w_{0\bk,\bar n} = {\sum_n}^\prime \alpha_{n}w_{n\bk} \,,\label{132}
\end{eqnarray}
where as before, the prime on the summation indicates that the $n=0$ 
term is excluded from the sum.
Inserting this expansion into (\ref{130}), and taking the inner product
with respect to $w_{n\bk}$, we obtain
\begin{eqnarray}
\frac{1}{2\bar{n}}N_{n0} + {\sum_{n'}}^\prime N_{nn'} \alpha_{n'}
-\mu_{0\bk,\bar{n}}\delta_{n0} = 0\,, \label{133a}
\end{eqnarray}
where we have noted that $\alpha^{*}_{n} = \eta_{n}\alpha_{n}$
as a result of the symmetry property (\ref{118a}).
Setting $n=0$ in (\ref{133a}), we find
\begin{eqnarray}
\mu_{0\bk,\bar{n}} = \frac{1}{2\bar{n}}N_{00} +
{\sum_{n}}^\prime N_{0n}\alpha_{n}. 
\label{mu-alpha}
\end{eqnarray}
The set of equations for $n\not=0$ has the solution
\begin{eqnarray}
\alpha_{n} =
-\frac{1}{2\bar{n}}{\sum_{n'}}^\prime (\widetilde N^{-1})_{nn'} 
N_{n'0}\,,
\label{134}
\end{eqnarray}
where $\widetilde N$ is the reduced matrix obtained by
deleting the first row and first column from $N$.
Thus we find that
\begin{eqnarray}
\mu_{0\bk,\bar{n}} = \frac{1}{2\bar{n}}\left (N_{00} -
{\sum_{nn'}}^\prime N_{0n} (\widetilde N^{-1})_{nn'} N_{n'0}\right ). 
\label{133b}
\end{eqnarray}
This is analogous to (\ref{25}) and reduces to it in the $\bk = 0$ limit
since the $M$ and $N$ matrices are then
the same (note that $w_{n{\bf 0}}$ are defined to be real).

\subsection{Determination of ${\mu}_{0\bk,i}$}

We follow a similar method to obtain $\mu_{0\bk,i}$.
Taking the derivative of $\hat{h}_{\bk}w_{0,\bk} = 0$ with respect
to $k_i$, we have
\begin{eqnarray}
&&\hskip -.1truein \Big[\frac{\hbar}{m}\Big(p_i + \hbar k_i\Big)
+ g\bar{n}\Big(w_{0\bk,i}^{*}w_{0\bk} + w_{0\bk}^{*}
w_{0\bk,i}\Big) - \mu_{0\bk,i}\Big]w_{0\bk}\nonumber\\
&&\hskip 2truein + \hat{h}_{\bk}w_{0\bk,i} = 0,  \label{135}
\end{eqnarray}
where $w_{0\bk,i} = \partial w_{0\bk}/\partial k_i$.
Noting the orthogonality of each vector component
$w_{0\bk,i}$ with $w_{0\bk}$, we have the following expansion
\begin{equation}
w_{0\bk,i} = {\sum_{n}}^\prime \beta_{in} w_{n\bk},  \label{136}
\end{equation}
where the expansion coefficients, $\beta_{in}$, define the
Cartesian components of a vector $\bm{\beta}_n$.
Inserting (\ref{136}) into (\ref{135}), and taking
the inner product with $w_{n\bk}$, we obtain
\begin{eqnarray}
\frac{\hbar}{m}(P_i)_{n0} + {\sum_{n'}}^\prime N_{nn'}\beta_{in'} -
\mu_{0\bk,i}\delta_{n0} = 0.
\label{137}
\end{eqnarray}
where, as before,  we have used $\beta^{*}_{in} = \eta_{n}\beta_{in}$. 
An expression for $\mu_{0\bk,i}$ can be found by setting $n=0$ in
(\ref{137}):
\beq
\mu_{0\bk,i} = \frac{\hbar}{m}(P_i)_{00} +
{\sum_{n}}^\prime N_{0n'} \beta_{in'}. \label{138}
\eeq
The set of equations for $n\not=0$ yield the solution vector,
\beq
\beta_{in}=-\frac{\hbar}{m}{\sum_{n'}}^\prime (\widetilde N^{-1})_{nn'}
(P_i)_{n'0}\,,
\label{139}
\eeq
and thus,
\beq
\mu_{0\bk,i} = \frac{\hbar}{m}\left ( (P_i)_{00} -
{\sum_{nn'}}^\prime N_{0n}(\widetilde N^{-1})_{nn'} (P_i)_{n'0} 
\right )\,.
\label{dmu_dk}
\eeq

\subsection{Excitation Energy}

These results will now be used to obtain an expression for the
excitation energy $E$ from (\ref{d^1}) and (\ref{quad}). 
Setting $n= 0$ in (\ref{d^1}) we have
\beq
{\sum_{n'}}^\prime N_{0n'} d_{n'}^{(1)} = E - {\hbar \over m}\bq\cdot
\P_{00} - N_{00}d_{0}^{(1)}
\label{exc_1}
\eeq
while for $n\ne 0$ we see that
\beq
d_{n}^{(1)} = -{\sum_{n'}}^\prime (\widetilde N^{-1})_{nn'} \left (
{\hbar \over m} \bq \cdot \P_{n'0}+ N_{n'0} d_0^{(1)} \right )
\eeq
The quantities on the right hand side of this equation are in fact 
related to the
expansion coefficients $\alpha_{n}$ in (\ref{134}) and $\beta_{in}$ in
(\ref{139}). We find the  simple relation
\begin{equation}
d_{n}^{(1)} = 2\bar n d_0^{(1)} \alpha_{n} + \bq \cdot \bm{\beta}_{n}. 
\label{140}
\end{equation}
Using this result in (\ref{exc_1}), we have
\bea
\hskip -.2truein \left ( N_{00} + 2\bar n {\sum_n}^\prime N_{0n} 
\alpha_n \right )
d_0^{(1)} &=& E - {\hbar \over m} \bq \cdot \P_{00} \nonumber \\
&-& {\sum_n}^\prime N_{0n} \bq \cdot \bm{\beta}_{n}
\eea
which with (\ref{mu-alpha}) and (\ref{138}) can be written as
\beq
2\bar n \mu_{0\bk,\bar n} d_0^{(1)} = E - \bq\cdot \nabla_\bk 
\mu_{0\bk}\,.
\label{d1_eq.1}
\eeq
We see that $d_0^{(1)}$ and $E$ are now related to each other through
physically meaningful and calculable  parameters.

We now substitute (\ref{140}) into (\ref{quad}) to obtain
\bea
&&\hskip -.25truein {\hbar ^2 q^2 \over 2m} + {\sum_n}^\prime {\hbar 
\over m} (\bq \cdot \P_{0n})(\bq\cdot \bm{\beta}_{n}) \nonumber \\
&&\hskip -.15truein = \left ( E - {\hbar \over m} \bq \cdot \P_{00} -
{\sum_n}^\prime {\hbar \over m} \bq \cdot \P_{0n} 2\bar n \alpha_n
\right ) d_0^{(1)}\,.
\label{d1_eq}
\eea
With (\ref{134}), the sum on the right hand side becomes
\bea
{\sum_n}^\prime {\hbar \over m} \P_{0n} 2\bar n \alpha_n &=& 
-{\sum_{nn'}}^\prime {\hbar \over m} \P_{0n} (\widetilde N^{-1})_{nn'}
N_{n'0}\nonumber \\
&=& 
-{\sum_{nn'}}^\prime N_{0n} (\tilde N^{-1})_{nn'}
{\hbar \over m} \P_{n'0}\nonumber \\
&=& 
{\sum_n}^\prime N_{0n} \bm{\beta}_n\,.
\eea
In going from the first to the second line, we have used the fact that
all the matrices have the transposition property $M_{n'n} = \eta_n
\eta_{n'} M_{nn'}$. With the expression for $\mu_{0\bk,i}$ in
(\ref{138}), (\ref{d1_eq})
thus becomes
\bea
\hskip -.25truein {\hbar ^2 q^2 \over 2m} &+& {\sum_n}^\prime {\hbar 
\over m} (\bq \cdot \P_{0n})(\bq\cdot \bm{\beta}_{n}) \nonumber \\
&=& \left ( E - \bq \cdot \nabla_\bk \mu_{0\bk}
\right ) d_0^{(1)} 
\label{d1_eq.2}
\eea
Eliminating $d_0^{(1)}$ from (\ref{d1_eq.1}) and (\ref{d1_eq.2}), we 
finally obtain
\begin{eqnarray}
&&\hskip -.35truein E =  \bq \cdot \nabla_\bk \mu_{0\bk} \nonumber \\
&& \hskip -.2truein + \sqrt{2\bar{n}\mu_{0\bk,\bar{n}}\left[\frac{\hbar^{2}q^{2}}{2m} 
 + \frac{\hbar}{m}{\sum_n}^\prime (\bq \cdot \P_{0n})(\bq \cdot 
 \bm{\beta}_{n})\right]}. 
 \label{151}
\end{eqnarray}
The sign of the square root is chosen to be positive to give a positive
excitation energy in the $\bk \to {\bf 0}$ limit.
The final quantity to interpret is the summation within the square
root.

\subsection{Effective Mass Tensor}

The square bracket in (\ref{151}) involves the tensor
\bea 
&&\hskip -.5truein \left ( {1\over m} \right )_{ij} \equiv {1\over m}\delta_{ij}+{2\over
m\hbar} {\sum_n}^\prime (P_i)_{0n}(\beta_j)_{n} \nonumber \\ 
&&  = {1\over m}\delta_{ij} - {2\over m^2} {\sum_{nn'}}^\prime 
(P_i)_{0n} (\widetilde N^{-1})_{nn'} (P_j)_{n'0}\,. 
\label{m_ij}
\eea
This expression is similar to the usual effective mass tensor defined on
the basis of $\bk\cdot \bp$ perturbation theory, although the structure
of the summation is different. To make contact with the $\bk\cdot \bp$
expression
we consider the $N_{nn'}$ matrix in the $\bk \to 0$ limit. Quite
generally, this matrix has the block structure
\beq
{\bf N}(\bk) = \left ( \begin{array}{cc} 
{\bf A_{++}} + {\bf B_{++}} + {\bf D_{+}} & {\bf A_{+-}}-{\bf B_{+-}} \\
{\bf A_{-+}} + {\bf B_{-+}} & {\bf A_{--}} - {\bf B_{--}} + {\bf D_{-}}
\end{array} \right )
\eeq
where the blocks are defined according to the parity index of the
various states. For example, the block in the upper-left corner 
contains matrix
elements between states with a positive parity index, $\eta = +1$. The
diagonal matrix ${\bf D}$ contains the energy eigenvalues
$\varepsilon_{n\bk}$ on its
diagonal. In the limit $\bk \to 0$, we have ${\bf B}(\bk = 0) = {\bf
A}(\bk = 0)$ and $A_{nn'} = 0$ if $\eta_n \ne \eta_{n'}$. Thus, in this
limit we have
\beq
{\bf N}(\bk=0) = \left ( \begin{array}{cc} 
2{\bf A_{++}} + {\bf D_+} & {\bf 0} \\
{\bf 0}  & {\bf D_-}\end{array} \right )
\eeq
that is, ${\bf N}(\bk =0)$ is block-diagonal, which of course is also
true of its inverse. Since, $P_i$ only connects states of {\it opposite}
parity in the $\bk = 0$ limit, we thus see that
\beq 
\lim_{\bk\to {\bf 0}}\left ( {1\over m} \right )_{ij} =
{1\over m}\delta_{ij} - {2\over m^2} {\sum_{n}}^\prime 
{(P_i)_{0n} (P_j)_{n0}\over \varepsilon_{n0}}\,. \nonumber
\label{m_ij_k=0}
\eeq
This is precisely the effective mass tensor obtained by means of
$\bk\cdot\bp$ perturbation theory~\cite{ashcroft76} as applied to the
Hamiltonian $\hat h_{\bf 0}(\bq)$ in (\ref{4}). The tensor defined in
(\ref{m_ij}) is a generalized effective mass tensor in that it depends
on the presence of a superfluid flow ($\bk \ne {\bf 0}$). Also because 
of this, it is no longer diagonal despite the cubic symmetry of the
optical lattice. 

To complete the identification of $(m^{-1})_{ij}$, we consider
variations of the GP equation with respect to the condensate wave
vector $\bk$. The second derivative of $\hat h_\bk w_{0\bk} = 0$ yields 
the equation
\beq
\hat h_{\bk,ij} w_{0\bk} + \hat h_{\bk,i} w_{0\bk,j} + \hat h_{\bk,j}
w_{0\bk,i} + \hat h_\bk w_{0\bk,ij} = 0\,.
\label{h_ij}
\eeq
Here,
\bea
\hat h_{\bk,i} &=& {\hbar \over m} (p_i+\hbar k_i) + g n_{c,i} -
\mu_{0\bk,i} \,,\\
\hat h_{\bk,ij} &=& {\hbar^2 \over m} \delta_{ij} + g n_{c,ij} -
\mu_{0\bk,ij} \,.
\eea
The inner product of (\ref{h_ij}) with $w_{0\bk}$ gives
\beq
{1\over \Omega}\int_\Omega w_{0\bk}^* \left ( \hat h_{\bk,ij} w_{0\bk}
+ \hat h_{\bk,i} w_{0\bk,j} + \hat h_{\bk,j} w_{0\bk,i} \right ) d^3r =
0\,.
\eeq
The first integral is
\beq
{1\over \Omega} \int_\Omega w_{0\bk}^* \hat h_{\bk,ij} w_{0\bk} d^3r = {\hbar^2 \over m}
\delta_{ij} - \mu_{0\bk,ij} + {g\over \bar n\Omega} \int_\Omega n_c 
n_{c,ij}\, d^3r
\eeq
while the integral of the next two terms gives the result
\bea
&&{1\over \Omega}\int_\Omega w_{0\bk}^* (\hat h_{\bk,i} w_{0\bk,j} + \hat h_{\bk,j}
w_{0\bk,i})d^3r \nonumber \\
&&\hskip .75truein = {\hbar \over m}
{\sum_n}^\prime [(P_i)_{0n} (\beta_j)_n + (P_j)_{0n} (\beta_i)_n] \nonumber \\
&&\hskip 1truein + {g\over \bar n\Omega} \int_\Omega n_{c,i} n_{c,j}
d^3r\,.
\eea
We thus find 
\beq
\mu_{0\bk,ij} = {\hbar^2 \over m} \delta_{ij} + {2\hbar \over m}
{\sum_n}^\prime (P_i)_{0n} (\beta_j)_n + {g\over 2 \bar n\Omega} 
\int_\Omega (n_c^2)_{,ij} d^3r\,.
\eeq
With this result we see that the effective mass tensor defined in
(\ref{m_ij}) can be expressed as
\beq
\left ( {1\over m} \right )_{ij} = {1\over \hbar^2} {\partial^2 \over 
\partial
k_i \partial k_j} \left ( \mu_{0\bk} - {g\over 2\bar n\Omega
} \int_\Omega n_c^2 \,d^3r \right )\,.
\label{m_ij.2}
\eeq

\subsection{Relation to Energy Density}

This last result can be related to the total energy
(\ref{E_tot}) in the state $\Phi_{0\bk}$.
Defining the energy per particle as $E_{\rm tot} \equiv
N\tilde \epsilon(\bar n, \bk)$, we have
\bea
\tilde \epsilon(\bar n, \bk) &=& {1\over \Omega} \int_\Omega w_{0\bk}^*
\left ( -{\hbar^2 \over 2m} (\nabla + i\bk)^2 + V_{\rm opt}
\right ) w_{0\bk} d^3r \nonumber \\
&&+ {g\bar n\over 2\Omega} \int_\Omega |w_{0\bk}|^4 d^3r
\eea
Comparing this with $\mu_{0\bk}$, we see that
\beq
\tilde \epsilon(\bar n, \bk) = \mu_{0\bk} - {g\bar n\over 2\Omega}
\int_\Omega |w_{0\bk}|^4 d^3r
\label{e/part}
\eeq
which is the expression in brackets in (\ref{m_ij.2}). 
Thus the effective mass tensor is given by 
\beq
\left ( {1\over m} \right )_{ij} = {1\over \hbar^2} {\partial^2 \tilde
\epsilon \over \partial k_i \partial k_j} 
\,.
\label{m_ij.3}
\eeq
For a cubic lattice, $\mu_{0\bk}$ has an expansion of the form 
$\mu_{0\bk} = \mu_{0{\bf 0}} +\hbar^2k^2/2m_\mu + \cdots$. Similarly,
$\tilde\epsilon(\bar n, \bk) = \tilde\epsilon(\bar n,0) + \hbar^2 k^2
/2 m_\epsilon +\cdots$. However, as proved by (\ref{m_ij_k=0}), the
parameter $m_\epsilon$ is in fact the band mass $m_0$ defined by the
Hamiltonian (\ref{4}). In other words, the correct effective mass
parameter can be extracted without solving the GP
equation for $w_{0\bk}$ (with $\bk \ne {\bf 0}$) self-consistently.
We note in passing that direct differentiation
of (\ref{e/part}) establishes the relation $\langle {\bf j}_s \rangle 
= \nabla_\bk (\bar n \tilde \epsilon)/\hbar$~\cite{diakonov02}.

With these results, the phonon energy
(\ref{151}) can be written in a compact form. Defining 
the mean energy density as $e \equiv \bar n \tilde \epsilon$, the 
Bogoliubov excitation energy at long wavelengths is given by
\beq
E = e_{,\bar n i}q_i + \sqrt{ e_{,\bar n \bar n} e_{,ij} q_i q_j}
\label{disp_rel}
\eeq
where we use a repeated summation convention on the Cartesian indices
$i$ and $j$.
This is precisely the expression given by Machholm {\it et
al.}~\cite{machholm03}  who
argued that the dynamics of the system at long wavelengths could be
based on a hydrodynamic analysis. Since their approach arrives at
(\ref{disp_rel}) in a more economical fashion, it is useful to 
summarize the essential assumptions on which it is based.

The central assumption is the existence of an average phase fluctuation,
$\langle \theta(\br,t) \rangle$, that varies slowly in space 
and time.  Expanding this average phase as 
\bea
\langle \theta(\br + \Delta \br, t+\Delta t) \rangle =  \langle
\theta(\br, t) \rangle &+& \nabla \langle \theta(\br, t) 
\rangle \cdot \Delta \br \nonumber \\
&& \hskip -.75truein +{\partial \langle \theta(\br, t) \rangle 
\over \partial t}\Delta t + \cdots\,,
\eea
one identifies $\nabla \langle \theta \rangle$ with 
the local wave vector, $ \langle \bk \rangle$, and $\partial 
\langle 
\theta \rangle/\partial t$ with $- \langle \mu \rangle 
/\hbar$ where $\langle \mu \rangle$ is the local
chemical potential. The equation of motion for the local wave vector is
thus
\beq
\hbar {\partial \langle \bk \rangle \over \partial t} = 
- \nabla \langle \mu \rangle\,.
\eeq
The second hydrodynamic equation is the continuity
equation
\beq
{\partial \langle n \rangle \over \partial t} + \nabla \cdot \langle
{\bf j}_s \rangle = 0\,,
\eeq
where $\langle {\bf j}_s \rangle$ is the local current density.
The current density and chemical potential
are then assumed to be given by the usual expressions for a
uniform system, namely,
\beq
\langle {\bf j}_s \rangle = {1\over \hbar} \nabla_{\langle \bk \rangle}
e\,,\qquad \langle \mu \rangle = {\partial e \over \partial \langle n
\rangle}
\eeq
where $e(\langle n \rangle, \langle \bk \rangle)$ is the average 
energy density for a uniform optical lattice,
viewed as a function of the local density
$\langle n \rangle$ and wave vector $\langle \bk
\rangle$~\cite{footnote2}. By expanding
the variables as $\langle n \rangle = \bar n + \delta n$ and $\langle
\bk \rangle = \bk + \delta \bk$, one obtains a pair of equations
for the fluctuations which admits wavelike solutions with frequency
$\omega=E/\hbar$ and wave vector $\bq$. The dispersion relation found is
identical to (\ref{disp_rel}).

It is clear that the assumptions made in the hydrodynamic approach are
completely justified. The average energy density $e$ is the fundamental
quantity determining the excitation energy at long wavelengths, as
confirmed by our systematic $q$-expansion. The additional information
provided by the expansion technique are the perturbative expressions 
for $\partial
\mu_{0\bk}/\partial \bar n$, $\nabla_\bk \mu_{0\bk}$ and $(1/m)_{ij}$ as
given by (\ref{133b}), (\ref{dmu_dk}) and (\ref{m_ij}), respectively.

\subsection{Discussion}

For small $\bk$, $e(\bar n,\bk) \simeq e(\bar n,{\bf 0}) + \bar n\hbar^2
k^2/2m_0 + \cdots$, and the Bogoliubov excitation energy is 
\beq
E/\hbar  \simeq \hbar \bq\cdot \bk {\partial \over \partial \bar n} \left (
{\bar n \over m_0} \right ) + sq
\eeq
where $s$ is the sound speed for the condensate at rest. This 
result was given previously by Kr\"amer {\it et al.}~\cite{kramer03}. 
The energy first becomes negative when the superfluid flow satisfies
$k>m_\mu s/\hbar$, where $ 1/m_\mu = \partial(\bar n/m_0)/\partial \bar
n$, which defines the Landau criterion for energetic instability at long
wavelengths in an optical lattice. The region of energetic instability
was mapped out for arbitrary $q$ by Wu and Niu~\cite{wu01,wu03} and 
Machholm {\it et al.}~\cite{machholm03}. 
In this region the energy of the superfluid state
is no longer a local maximum. As a result, transitions to lower energy
states can occur spontaneously provided a means of conserving energy and
quasimomentum is available.

The excitation energy given by (\ref{disp_rel}) becomes imaginary when 
the argument of the
square root is negative. This signals a dynamic instability whereby the
amplitude of the condensate fluctuation grows (or decays) in time. Of
the two factors in the square root, $e_{,ij}$, or equivalently the
effective mass tensor, $(1/m)_{ij}$, is the most physically relevant.

It is instructive to examine the latter in the weak potential limit. We
consider for simplicity the one-dimensional situation discussed in
Sec.~\ref{weak_potential}. Repeating the perturbative analysis in
Sec.~\ref{weak_potential} for the case $\bk \ne {\bf 0}$, we find
\beq
\tilde \epsilon(\bar n,k) = {1\over 2} g \bar n + \varepsilon^{(0)}_k - 
{V_0^2 \over 2\left (\varepsilon_G^{(0)}+2g\bar n
-4\varepsilon_k^{(0)}\right )} + \cdots
\label{tilde_eps}
\eeq
We note that this expression becomes singular at a wave vector $k_c$
satisfying $\varepsilon_G^{(0)} + 2g\bar n -4\varepsilon_{k_c}^{(0)} = 
0$, which gives
\beq
k_c = \sqrt{k_0^2 + G^2/4}
\label{k_c}
\eeq
where $k_0 = ms_0/\hbar$. The singularity is indicating the breakdown of
nondegenerate perturbation theory, but provided $k$ is not too close to
$k_c$, we can use (\ref{tilde_eps}) to evaluate the effective mass. To
second order in $V_0$ we have
\bea
\hskip -.25truein {1\over m_0(k)} &=& {1\over m} \left [ 1 - {2 V_0^2 \over
\left (\varepsilon_G^{(0)}+2g\bar n -4\varepsilon_k^{(0)}\right )^2} 
\right .\nonumber \\
&& \hskip .4truein - \left .{ 32 \varepsilon_k^{(0)}
V_0^2 \over \left (\varepsilon_G^{(0)}+2g\bar n -4\varepsilon_k^{(0)}\right )^3} 
+ \cdots \right ]
\eea
At $\bk = {\bf 0}$ we recover the result (\ref{meff}) found in
Sec.~\ref{weak_potential}. For $m_0^{-1}$ to go to zero, $k$ must be
close to $k_c$. With $\Delta k = k_c -k$, we find
\beq
{\Delta k \over k_c} \simeq \left ( {V_0 \over \varepsilon_G^0 + 2g\bar
n} \right ) ^{2/3}\,,
\eeq
that is, the wave vector $k$ approaches $k_c$ as $V_0 \to 0$. We note
that at $k=k_c-\Delta k$, the perturbative correction to the energy in 
(\ref{tilde_eps}) is still small so that the perturbation theory
estimate of where $m_0^{-1}$ goes to zero is reasonable. We thus
expect a dynamic instability to set in when $k \simeq k_c$ in the weak
potential limit.

This condition for the dynamical instability is the $q =  0$
limit of the result given in Refs.~\cite{wu01} and \cite{machholm03}. 
The Bogoliubov excitations 
of wave vector $q$ in an homogeneous gas with current $\langle j_s 
\rangle = \bar n \hbar k/m$ have the energies
\beq
E_\pm(q) = {\hbar^2 kq \over m} \pm \sqrt{{\bar n g \hbar^2 q^2\over m}
+ {\hbar^4 q^4\over 4m^2}}\,.
\label{homog_gas}
\eeq
We follow Wu and Niu~\cite{wu01} in referring to the modes with the
plus (minus) sign as phonons (anti-phonons). The former correspond to
physical excitations in that their normalization is given by
(\ref{uv_ortho}). The effect of an optical potential is to couple an
anti-phonon mode with wave vector $q$ to a phonon mode with wave vector
$q-G$. The condition that  $E_-(q)=E_+(q-G)$ implies that the two modes
are resonantly coupled and gives the critical wave vector
\beq
k_c= {1\over G} \left ( \sqrt{k_0^2q^2 +{1\over 4}q^4}+\sqrt{k_0^2(G-q)^2
+{1\over 4}(G-q)^4}\right )\,.
\label{k_c_q}
\eeq
For $q=0$, this gives the critical wave vector in (\ref{k_c}).
The expression in (\ref{k_c_q}) was shown in \cite{wu01} to account for 
the boundary of the dynamically unstable region in the weak potential 
limit. In fact, it can be shown by means of degenerate perturbation 
theory (Appendix~\ref{appendixB}) that imposing a weak
optical potential indeed gives rise to complex Bogoliubov
eigenvalues. 

Alternatively, the condition  $E_-(q)=E_+(q-G)$ can be
written as $E_+(-q)+E_+(q-G)=0$. This was interpreted by Machholm {\it
et al.} as a Landau criterion for the emission of two phonon 
excitations with zero total energy.  Although this physical
interpretation is appealing, it is not clear how it can be used to 
actually determine the rate at which the excitations are being produced,
short of performing the perturbation analysis carried out in 
Appendix~\ref{appendixB} in terms of {\it phonon and anti-phonon} 
modes.

We thus see that the phonon-anti-phonon resonance condition, or
alternatively the
two-phonon Landau criterion, is consistent with the effective mass 
condition for a dynamical instability in the $q\to 0$ limit. A similar
statement can be made in the weak coupling limit ($g \to 0$). 
Wu and Niu~\cite{wu01} noted from their numerical analysis that one
boundary of the dynamically unstable region is given by the condition
$\varepsilon_0(q+k)-\varepsilon_0(k) =
\varepsilon_0(k)-\varepsilon_0(k-q)$ where $\varepsilon_0(k)$ 
is the band energy for the optical potential by itself.
In the small-$q$ limit, this condition becomes
\begin{displaymath}
{\partial^2 \varepsilon_0(k) \over \partial k^2} = 0\,.
\end{displaymath}
Thus the onset of dynamical instability at $q=0$ in the weak
coupling limit is again given by the point at which the inverse
effective mass goes to zero. 

%% file: conclusions.tex
\section{Conclusions}

We have studied the long wavelength phonon excitations in a three
dimensional optical lattice. By making use of a systematic expansion of
the Bogoliubov equations in
terms of the phonon wave vector $\bq$, we obtain the phonon dispersion
in the long wavelength limit. Our result (\ref{26}) for the 
current-free state defines the sound speed in terms of the effective
mass $m_0$ and variations of the chemical potential with $\bar n$ and 
agrees with the result given by Menotti {\it et al.}~\cite{menotti02}.
The effective mass is defined quite generally in terms of the energy per
particle, $\tilde \epsilon(\bar n,\bk)$, but can also be calculated
using the current-free GP Hamiltonian in the $\bk \to {\bf 0}$ limit. 
We present analytic expressions
for the sound speed in the Thomas-Fermi, weak potential, weak coupling
and tight-binding limits.

For the current-carrying case, we rederive the dispersion relation
obtained by means of a hydrodynamic analysis~\cite{machholm03} (see also
\cite{kramer02}). Our approach confirms that the dynamics at long
wavelengths is defined by the local energy density $e(\langle n \rangle,
\langle \bk \rangle)$ viewed as a function of
the slowly varying local density, $\langle n(\br) \rangle$, and local
condensate wave vector, $\langle \bk(\br) \rangle$. At long wavelengths,
dynamical instabilities arise at the point where the generalized
effective mass tensor has a vanishing eigenvalue.

%% file: appendix1.tex
\begin{appendix}

\section{Reflection Symmetry of Bloch Functions}
\label{appendixA}
In this Appendix we give a proof of the symmetry property (\ref{118a}) 
used
throughout our analysis. We do this for the one-dimensional case for
which the wave function is a solution of
\beq
-{\hbar^2 \over 2m} {d^2\psi \over dx^2} + V(x) \psi(x) = E \psi(x)\,,
\label{Sch_eq}
\eeq
where the potential is periodic, $V(x+d) = V(x)$, and is assumed to have
inversion symmetry, $V(-x) = V(x)$. In the context of the GP equation,
$V(x) = V_{\rm opt}(x) + gn_c(x)$ and the inversion property is valid if
the condensate density also satisfies $n_c(-x) = n_c(x)$. This is
ensured if the wave function has the property we wish to prove.

We seek solutions of the Bloch form, $\psi(x+d) = e^{ikd} \psi(x)$.
Due to the inversion symmetry, the linearly independent
solutions of (\ref{Sch_eq}) can be chosen to be even ($\psi_e$) or odd ($\psi_o$)
functions of $x$ and $\psi(x)$ can be expressed as the linear
combination
\beq
\psi(x) = a \psi_e(x) + b \psi_o(x)\,.
\eeq
The periodic part of the Bloch function is then
\beq
w_k(x) = e^{-ikx}\left ( a \psi_e(x) + b \psi_o(x) \right )\,.
\eeq
The two independent solutions at energy $E$ are chosen to have the
normalization
\beq
{1\over d} \int_{-d/2}^{d/2} |\psi_{e,o}(x)|^2\, dx = 1\,.
\eeq
Imposing the Bloch condition, we obtain the relation
\beq
{\psi_e^\prime \over \psi_e} = - {\psi_o^\prime \over
\psi_o}\tan^2\left ( {kd\over 2 } \right )
\eeq
where all the functions are evaluated at $x=d/2$. Since $\psi_e$ and
$\psi_o$ are functions of the energy, $E$, this equation determines the
band energy $E_k$. Clearly $E_{-k} = E_k$. If $E_0$ is the band energy
at the zone centre ($k=0$), we must have either $\psi'_e(d/2;E_0)=0$ or
$\psi_o(d/2;E_0)=0$. The former defines what we shall refer to as 
an even-parity band, while the latter defines
an odd-parity band. The small-$k$ behaviour of $E$ is thus
readily obtained from these properties. For example, for an odd-parity
band we have
\beq
\psi_o(d/2;E) = \dot \psi_o (E-E_0) + \cdots
\eeq
where $\dot \psi_o \equiv d\psi_o(d/2;E)/dE|_{E=E_0}$. We then have
\beq
E_k = E_0 - {\psi_o' \psi_e' \over \dot \psi_o \psi_e}\left ( 
{dk\over 2}\right )^2+\cdots
\eeq
The coefficient of $k^2$ defines the effective mass of the band. A
similar result applies in the case of the even-parity bands.

Once the energy eigenvalue for a given $k$ is known, the coefficients
$a$ and $b$ are related by
\beq
{b\over a} = i{\psi_e \over \psi_o} \tan\left ( {kd\over 2}\right )\,.
\eeq
For a given band, $n$, the ratio $b/a$ is a continuous function of $k$.
At $k=0$ we choose $w_{k=0}(x)$ to be real and assume that it is a
parity eigenstate. In this situation, we must have either $b(k=0)=0$
(even-parity bands) or $a(k=0)=0$ (odd-parity bands).

The normalization of $w_k$ leads to the expressions
\beq
|a|^2={1 \over 1+ \lambda^2}\,,\qquad
|b|^2={\lambda^2 \over 1+ \lambda^2}
\eeq
where
\beq
\lambda(k) = {\psi_e\over \psi_o}\tan\left ( {kd\over 2}\right )\,.
\eeq
For an even-parity band $\lambda \to 0$ as $k \to 0$, so that $a(k) \to
1$ and $b(k) \to 0$. In this case,
\beq
w_k(x) = e^{-ikx} \left ( {\psi_e(x) \over \sqrt{1+\lambda^2}} +
{i\lambda \psi_o(x) \over \sqrt{1+\lambda^2}} \right )\,.
\eeq
Since $\lambda(-k) = - \lambda(k)$, we see that $w_{-k}(-x) = w_k(x)$.
On the other hand, for an odd-parity band, $\lambda(k) \to \infty$ as $k
\to 0$, and $b(k) \to 1$. As a result, we have
\beq
w_k(x) = e^{-ikx} \left({-i\, {\rm sgn}(\lambda) \psi_e(x) \over \sqrt{1+\lambda^2}} +
{|\lambda| \psi_o(x) \over \sqrt{1+\lambda^2}} \right )\,,
\eeq
which implies $w_{-k}(-x) = -w_k(x)$. We have thus shown that the Bloch
functions have the property
\beq
w_{-k}(-x) = \pm w_k(x)
\eeq
where the positive (negative) sign corresponds to the even (odd) parity
bands. Together with the conjugation property $w_k^*(x) = w_{-k}(x)$,
we have $w_k^*(-x) = \pm w_k(x)$.

For an even-parity band, the real and imaginary parts of $w_k$ are
\bea
\Re w_{k}(x) &=& {1\over \sqrt{1+\lambda^2}}\left ( \cos(kx)\psi_e(x) +
\lambda \sin(kx) \psi_o(x) \right )\nonumber \\
\Im w_{k}(x) &=& {1\over \sqrt{1+\lambda^2}}\left (-\sin(kx) \psi_e(x) +
\lambda \cos(kx) \psi_o(x) \right )\,.\nonumber 
\eea
Thus the real part is an even function of $x$ with the
property $d \Re w_k/dx|_{\pm d/2} = 0$, while the imaginary part is odd
and $\Im w_k(\pm d/2) = 0$. The opposite is true of an odd-parity band.
One can show for an arbitrary $k$ in
the lowest band that there is no net change in the phase 
$\theta_k(x) = \tan^{-1}(\Im
w_k(x)/\Re w_k(x))$ as $x$ varies between $-d/2$ and
$d/2$. We make use of this result in Sec.~\ref{basic}.

The method described above cannot be used in three dimensions, but
perturbation
theory allows one to infer the same symmetry property. We write the
Schr\"odinger equation for the Bloch function $w_\bk(\br)$ as
\beq 
(\hat h_0 + \delta V) w_\bk(\br)
\eeq
 where $\hat h_0 = -(\hbar^2/2m)\nabla^2 + \hbar^2 k^2/2m + V(\br)$ 
 and $\delta V = (\hbar \bk/m)\cdot \nabla$. The eigenfunctions of
 $\hat h_0$, $w_n(\br)$, with eigenenergies $E_n$, are chosen to be 
 parity eigenstates. The state $w_\bk$ to first order in $\delta
 V$ is
 \beq
 w_{\bk} = w_n + {\sum_{n'}}' {(\hbar/ m) \bk \cdot \P_{n'n}
 \over E_n - E_{n'}} w_{n'}
 \eeq
 where $\P_{n'n}$ is the $\bk =0$ momentum matrix element defined in
 (\ref{relations}). Since this matrix element only couples states with 
 opposite parity, we see that $w_{-\bk}(-\br) = \pm
 w_{\bk}(\br)$, depending on the parity of the state $n$. 
 Thus, the Bloch state exhibits the symmetry property to lowest order in
 $\bk$. It is evident that this argument can be extended to higher
 orders in perturbation theory.

\section{Dynamic Instability in the Weak Potential Limit}
\label{appendixB}

As pointed out by Wu and Niu~\cite{wu01}, the boundary of the
dynamically unstable region in the weak potential limit is given by the
condition $E_-(q) = E_+(q-G)$ where $E_\pm(q)$ is given by
(\ref{homog_gas}). These energies are the eigenvalues of
the Bogoliubov equations 
\beq
\hat B_0 \left ( \begin{array}{c} u_\pm \\ v_\pm \end{array} \right ) =
E_\pm  \left (
\begin{array}{c} u_\pm \\ -v_\pm \end{array} \right )
\eeq
with
\beq
\hat B_0 = \left ( \begin{array}{cc} -{\textstyle{\hbar^2}\over
\textstyle{2m}}{\textstyle{d^2}\over \textstyle{dz^2}} +
g\bar n - \varepsilon_k^{(0)} & -g\bar n e^{2ikz} \\
-g\bar n e^{-2ikz} & -{\textstyle{\hbar^2}\over
\textstyle{2m}}{\textstyle{d^2}\over \textstyle{dz^2}} + 
g\bar n - \varepsilon_k^{(0)} \end{array} \right )
\eeq
The phonon mode of interest is
\beq
\left ( \begin{array}{c} u_+ \\ v_+ \end{array} \right ) = \left (
\begin{array}{c} a_+ e^{i(q-G+k)z} \\ b_+ e^{i(q-G-k)z} \end{array} 
\right )
\eeq
with $a_+ = \sqrt{(\tilde \varepsilon_{q-G}+\tilde E_+) /2\tilde E_+}$
and
$b_+ = \sqrt{(\tilde \varepsilon_{q-G}-\tilde E_+) /2\tilde E_+}$ where
$\tilde \varepsilon_{q-G} = \varepsilon_{q-G}^{(0)} + g\bar n$ and $\tilde E_+ 
= E_+(q-G) - \hbar^2 k(q-G)/m$. The normalization of the mode
is $a_+^2 - b_+^2 = 1$. The anti-phonon mode which is coupled to the
phonon mode by a weak optical potential, $V_{\rm opt} = V_0 \cos(Gz)$,
is 
\beq
\left ( \begin{array}{c} u_- \\ v_- \end{array} \right ) = \left (
\begin{array}{c} a_- e^{i(q+k)z} \\ b_- e^{i(q-k)z} \end{array} \right )
\eeq
with $a_-=\sqrt{(\tilde \varepsilon_q - \tilde E_-) /2\tilde E_-}$ and
$b_-=\sqrt{(\tilde \varepsilon_q + \tilde E_-) /2\tilde E_-}$, where
$\tilde E_- = E_+(q) - \hbar^2 kq/m$.\hfil  We note that in this case the 
mode has normalization $a_-^2 - b_-^2 = -1$. 

The degeneracy of the phonon and anti-phonon modes
($E_-(q) = E_+(q-G) \equiv E_0$) suggests that we 
seek a solution of the Bogoliubov equations 
\beq
\hat B \left ( \begin{array}{c} u \\ v \end{array} \right ) =
E  \left (
\begin{array}{c} u \\ -v \end{array} \right )
\label{bog_eq}
\eeq
in the form
\beq
\left ( \begin{array}{c} u \\ v \end{array} \right ) = A \left (
\begin{array}{c} u_+ \\ v_+ \end{array} \right ) + B \left (
\begin{array}{c} u_- \\ v_- \end{array} \right )\,.
\eeq
Expanding the operator $\hat B$ to first order in the optical
potential, we have $\hat B = \hat B_0 + \hat B_1$ 
with
\beq
\hat B_1 = \left ( \begin{array}{cc} V_{\rm opt} + 2g\bar n(w_1
+w_1^*) & -2g\bar n e^{2ikx} w_1 \\
-2g\bar n e^{-2ikx} w_1^* & V_{\rm opt} + 2g\bar n(w_1 +w_1^*)
\end{array} \right )
\eeq
Here we have written the condensate wave function as $\Phi_k(x) =
\sqrt{\bar n} e^{ikx} ( 1 + w_1 + \cdots)$ where the first order
correction is $w_1(x) = \alpha_+ e^{iGx} + \alpha_- e^{-iGx}$ with
\beq
\alpha_\pm = - {\Big [ \varepsilon_G^{(0)} \mp \sqrt{4 \varepsilon_G^{(0)}
\varepsilon_k^{(0)}}\Big ]V_0 \over 2\left [ \big ( \varepsilon_G^{(0)} \big )^2 + 2g\bar n
\varepsilon_G^{(0)} -4 \varepsilon_G^{(0)}\varepsilon_k^{(0)}\right  ] }
\,.
\eeq
Taking the inner product of (\ref{bog_eq}) with $(u_+^*\,\, v_+^*)$ and
$(u_-^*\,\, v_-^*)$, and noting the different normalizations of the two
modes, we obtain the matrix equation
\beq
\left ( \begin{array}{cc} E_0 - E & \Delta \\ \Delta & -E_0 +E
\end{array} \right ) \left ( \begin{array}{c} A \\ B \end{array} \right
) = 0\,,
\eeq
where the real coupling parameter $\Delta$ is given by
\bea
&&\Delta = \left ( {V_0 \over 2} + 2g\bar n (\alpha_+ + \alpha_-) 
\right ) (a_+ a_- + b_+ b_-) \nonumber \\
&&\hskip .5truein -2g\bar n(\alpha_- a_+ b_- + \alpha_+ a_- b_+)\,.
\eea
A nontrivial solution to the matrix equation is obtained if
\beq
E = E_0 \pm i|\Delta|\,.
\eeq
Thus, the line in the $k$-$q$ plane defined by (\ref{k_c_q}) lies 
{\it within} the region of dynamical instability when $V_0$ is finite.
As emphasized by Wu and Niu~\cite{wu01}, the dynamical instability in
the weak potential limit arises from a resonant coupling between
phonon and anti-phonon modes.

\end{appendix}

%% file: sound_speed.bbl
\begin{thebibliography}{99}

\bibitem{cataliotti01} F.S. Cataliotti, S. Burger, C. Fort, P.
Maddaloni, F. Minardi, A Trombettoni, A. Smerzi and M. Inguscio, Science
{\bf 293}, 843 (2001).
\bibitem{anderson98} B.P. Anderson and M.A. Kasevich, Science
{\bf 282}, 1686 (1998).
\bibitem{morsch01} O. Morsch, J.H. M\"uller, M. Cristiani, D. Ciampini
and E. Arimondo, Phys. Rev. Lett. {\bf 87}, 140402 (2001).
\bibitem{burger01} S. Burger, F.S. Cataliotti, C. Fort, F. Minardi, M.
Inguscio, M.L. Chiofalo and M.P. Tosi, Phys. Rev. Lett. {\bf 86}, 4447
(2001).
\bibitem{berg-sorensen98} K. Berg-S\o rensen and K. M\o lmer, Phys. 
Rev. A {\bf 58}, 1480 (1998).
\bibitem{wu01} B. Wu and Q. Niu, Phys. Rev. A {\bf 64}, 061603(R) 
(2001).
\bibitem{machholm03} M. Machholm, C.J. Pethick, and H. Smith, Phys. Rev.
A {\bf 67}, 053613 (2003).
\bibitem{bronski01} J.C. Bronski, L.D. Carr, B. Deconinck, J.N. Kutz and
K. Promislow, Phys. Rev. E {\bf 63}, 036612 (2001).
\bibitem{konotop02} V.V. Konotop and M. Salerno, Phys. Rev. A {\bf 65},
021602 (2002).
\bibitem{javanainen99} J. Javanainen, Phys. Rev. A {\bf 60}, 4902  
(1999).
\bibitem{chiofalo00} M.L. Chiofalo, M. Polini and M.P. Tosi, Euro. Phys.
J. D {\bf 11}, 371 (2000).
\bibitem{smerzi02} A. Smerzi, A. Trombettoni, P.G. Kevrekidis and A.R.
Bishop, Phys. Rev. Lett. {\bf 89}, 170402 (2002).
\bibitem{kramer02} M. Kr\"{a}mer, L. Pitaevskii, and S. Stringari, 
Phys. Rev. Lett., {\bf 88}, 180404 (2002).
\bibitem{callaway91} J. Callaway, {\it Quantum Theory of the Solid
State} (Academic Press, New York, 1991).
\bibitem{pu03} H. Pu, L.O. Baksmaty, W. Zhang, N.P. Bigelow and P.
Meystre, Phys. Rev. A {\bf 67}, 043605 (2003).
\bibitem{bender78} C.M. Bender and S.A. Orszag, {\it Advanced
Mathematical Methods for Scientists and Engineers} (McGraw-Hill, New
York, 1978).
\bibitem{menotti02} C. Menotti, M. Kr\"amer, L. Pitaevskii and S.
Stringari, cond-mat/0212299.
\bibitem{kramer03} M. Kr\"amer, C. Menotti, L. Pitaevskii and S.
Stringari, cond-mat/0305300.
\bibitem{machholm03b} M. Machholm, A. Nicolin, C.J. Pethick and H.
Smith, cont-mat/0307183.
\bibitem{diakonov02} D. Diakonov, L.M. Jensen, C.J. Pethick and H.
Smith, Phys. Rev. A {\bf 66}, 013604 (2002).
\bibitem{hutchinson97} D.A. W. Hutchinson, E. Zaremba and A. Griffin,
Phys. Rev. Lett. {\bf 78}, 1842 (1997).
\bibitem{zaremba98} E. Zaremba, Phys. Rev. A {\bf 57}, 518 (1998).
\bibitem{jackson98}A.D. Jackson, G.M. Kavoulakis, and C.J. Pethick, 
Phys. Rev. A  {\bf 58}, 2417 (1998).
\bibitem{trombettoni01} A. Trombettoni and A. Smerzi, Phys. Rev. Lett.
{\bf 86}, 2353 (2001).
\bibitem{footnote3} More precisely, we fix the direction of the phonon
wave vector and order the expansion according to
powers of the wave vector magnitude $q$.
\bibitem{footnote2} Alternatively, the hydrodynamic equations can be
derived by taking variations of the action $S = \int d^3r\int dt \left 
\{ e(\langle n
\rangle, \langle \bk \rangle ) + \hbar \langle n \rangle \partial
\langle \theta \rangle/\partial t\right \}$ with $ \langle \bk \rangle =
\nabla \langle \theta \rangle$.
\bibitem{ashcroft76} N. W. Ashcroft and N. D. Mermin, {\it Solid
State Physics}, (Harcourt, 1976), pg. 765.
\bibitem{wu03} B. Wu and Q. Niu, cond-mat/0306411.
\bibitem{footnote_1} The other possibility of phonons with wave vectors
$q$ ($q > 0$) and $-G-q$ gives a critical wave vector larger than
(\ref{k_c}).
\end{thebibliography}
